\documentclass[twocolumn,tracking]{aastex702}
\usepackage[T1]{fontenc}
\usepackage{amsmath}
\usepackage{bm}
\usepackage{newtxmath}
\usepackage[dvipsnames]{xcolor}
\usepackage{graphicx}
\usepackage{physics}
\graphicspath{{./fig/}}
\newcommand{\ci}{\mathrm{i}} %
\defcitealias{schroer2025}{SCB}

\newcommand{\kmin}{{k_\mathrm{min}}}

\newcommand{\valf}{{\varv_\mathrm{A}}}
\newcommand{\vdr}{{\varv_\mathrm{dr}}}
\newcommand{\crrm}{{\mathrm{cr}}}
\newcommand{\NLLD}{\mathrm{NLLD}}
\newcommand{\vwa}{\varv_\mathrm{wave}}
\newcommand{\vbeat}{\varv_\mathrm{beat}}

\newboolean{draft}
\setboolean{draft}{false}

\DeclareMathOperator{\sign}{sign}

\begin{document}
\author[0000-0002-4683-8517]{Rouven Lemmerz}
\affiliation{Department of Physics, University of Wisconsin-Madison, Wisconsin 53706, USA}
\email{lemmerz@wisc.edu}

\author[0000-0001-9625-5929]{Mohamad Shalaby}
\affiliation{Perimeter Institute for Theoretical Physics, 31 Caroline St.~N., Waterloo, Ontario N2L 2Y5, Canada}
\affiliation{Waterloo Centre for Astrophysics, University of Waterloo, Waterloo, ON N2L 3G1, Canada}
\affiliation{Department of Physics and Astronomy, University of Waterloo, Waterloo, ON, N2L 3G1, Canada}
\affiliation{Horizon AstroPhysics Initiative (HAPI) Fellow}
\email{mshalaby@live.ca}

\author[0000-0003-4821-713X]{Ellen Zweibel}
\affiliation{Department of Astronomy, University of Wisconsin-Madison, Wisconsin 53706, USA}
\affiliation{Department of Physics, University of Wisconsin-Madison, Wisconsin 53706, USA}
\email{zweibel@astro.wisc.edu}
\author[0000-0002-7275-3998]{Christoph Pfrommer}
\affiliation{Leibniz-Institut f{\"u}r Astrophysik Potsdam (AIP), An der Sternwarte 16, 14482 Potsdam, Germany}
\email{cpfrommer@aip.de}

\author[0000-0002-7443-8377]{Timon Thomas}
\affiliation{Leibniz-Institut f{\"u}r Astrophysik Potsdam (AIP), An der Sternwarte 16, 14482 Potsdam, Germany}
\email{tthomas@aip.de}

\title{%
	The impact of nonlinear Landau damping on cosmic ray streaming and the role of external turbulence
}
	\shorttitle{Impact of Nonlinear Landau damping on CR streaming}
    
	\begin{abstract}
Cosmic rays (CRs) streaming faster than the Alfvén speed generate Alfvén waves that scatter them. The saturated wave amplitude is regulated by a damping mechanism. This damping controls the CR scattering rate and drift velocity, and thus the rate at which CRs transfer momentum and energy to the ambient plasma, which sets the strength of CR feedback. In the ionized interstellar medium, nonlinear Landau damping (NLLD) is commonly invoked as the dominant damping mechanism for Alfvén waves. We test this assumption with fluid-PIC simulations in which thermal ions and electrons are modeled as fluids with optional Landau closures, allowing us to compare otherwise identical runs with and without NLLD. When waves are self-generated by monoenergetic CRs, runs with and without Landau closures are indistinguishable: NLLD is inefficient and the CRs isotropize in the wave frame. Once high-amplitude Alfvén waves mimicking external turbulence are seeded, NLLD curbs wave growth and equalizes forward- and backward-propagating waves. We show that the saturated drift velocity, which falls below the Alfvén speed, can be accurately predicted from the wave spectra alone. If the CR distribution is undersampled, a numerical instability corrupts the largest resonant scales: unphysical, unresolved pitch-angle gradients resonantly generate backward-propagating waves, driving an inverse cascade that should be absent from purely streaming-driven waves. While power-law CR distributions may permit stronger NLLD, our results indicate that NLLD is unlikely to be an intrinsic saturation mechanism of the streaming instability; instead, its effectiveness hinges on the interplay of streaming-generated modes with those of external origin.

	\end{abstract}


	\section{Introduction}
	Cosmic rays (CRs) carry an energy density comparable to that of the thermal gas, the turbulence and the magnetic field in the interstellar medium (ISM) combined. 
	Being collisionless, they can exchange this energy with the gas through the electromagnetic fields of the Alfv\'en waves that scatter them \citep{zweibel2017,ruszkowskipfrommer2023}.
	CRs drifting faster than the Alfv\'en speed $\valf$ amplify these waves themselves \citep{wentzel1968HydromagneticWavesExcited,kulsrud1969,skilling1975CosmicRayStreamingI}: CRs are self-confined and their bulk drift is regulated close to $\valf$ for as long as the waves are free to grow.
	The amplitude at which the waves saturate is therefore set by the competition between this growth and whatever damps the waves or disrupts the instability, so that the damping or limiting process controls both how quickly CRs escape a galaxy and how hard they push on the gas on their way out.

	Which damping process dominates depends on a number of factors.
	Early considerations of CR self-confinement in the galaxy \citep{kulsrud1969} already considered the quenching of the streaming instability through ion-neutral damping in weakly ionized media, but a large fraction of the ISM is too ionized for these collisions to matter.
	In the ionized phases two candidates remain: shearing apart of resonant waves by an ambient turbulent cascade \citep{farmer2004WaveDampingMagnetohydrodynamic}, and nonlinear Landau damping (NLLD) \citep{hollweg1971NonlinearLandauDamping,lee1973DampingNonlinearWaveparticle}, a process in which two incompressible Alfv\'en waves nonlinearly generate a compressional ion-acoustic beat wave that is susceptible to collisionless Landau damping.
	Because the NLLD rate consequently grows with the wave energy and, unlike turbulent damping, requires no externally supplied cascade, the streaming modes could in principle damp themselves via NLLD. Whether they in fact do is the question of this paper.

	NLLD has been directly invoked as the damping mechanism for CR streaming in the ionized ISM \citep{volkCesarsky1982,miller1991} and has since become a standard ingredient of CR transport theory. It has been used to model features of the observed CR spectrum such as spectral breaks \citep{blasi2012} and to regulate CR-driven galactic winds \citep{wiener2017CosmicRaydrivenGalactic,thomas2026,sike2025}, and it has entered galactic simulation frameworks more broadly \citep{thomas2019CosmicrayHydrodynamicsAlfvenwave,armillotta2021,hopkins2021TestingPhysicalModels}. 
	Among the damping processes available in the ISM, NLLD is one of the weaker ones, and it operates predominantly in the halo \citep{xu2022,recchia2022}.
	This relative weakness is precisely what makes it important for understanding the escape of CRs from the galactic disk: the effective CR diffusion is dominated by the regions in which the streaming instability is damped the least, that is, by the regions in which NLLD sets the wave amplitude \citep{thomas2026}.
	NLLD also becomes more efficient in regions that have a high plasma-$\beta$, such as the intracluster medium, where NLLD has been proposed as a heating mechanism \citep{loewenstein1991,guo2008,wiener2013CosmicRayStreaming,pfrommer2013,ruszkowski2017}.
	 
	Observationally, the CR diffusion coefficient is constrained by the grammage that CRs accumulate before they escape the galaxy, as measured through secondary-to-primary ratios such as boron-to-carbon \citep[e.g.,][]{ptuskin2008NonlinearDiffusionCosmic,aloisio2015NonlinearCosmicRay,evoli2018}.
	Whether the scattering rate implied by these measurements is consistent with NLLD providing the dominant damping remains contested, reaching both negative \citep{hopkins2021TestingPhysicalModels,hopkins2022,kempski2022} and positive \citep{chernyshov2022,chernyshov2024} conclusions.
	
	Part of this disagreement, and differing results in the CR community more broadly, can be traced back to the different NLLD prescriptions in use.
	The damping rate derived by \citet{lee1973DampingNonlinearWaveparticle} is generally agreed upon -- and has been confirmed numerically for individual interactions to good accuracy \citep{holcombMicrophysicsGyroresonantStreaming2019,lemmerz2024CouplingMultifluidDynamics} -- but it describes the interaction of a single pair of Alfv\'en modes and is too detailed for use in coarse-grained models. 
	The published extensions to an Alfv\'en wave spectrum are ambiguous: they either sum the wave-wave interactions over all pairs of modes in the spectrum \citep[e.g.,][]{miller1991}, or retain only the wave-wave interactions that are local in $k$-space \citep[e.g., Chapter 11.5 of][]{kulsrud2004PlasmaPhysicsAstrophysics}, justified by the trapping of background particles, or model NLLD as a diffusion in $k$-space \citep[e.g.,][]{blasi2012,aloisio2013}, justified by ease of use.
	Because these prescriptions weight the wave spectrum differently, they predict significantly different damping rates for one and the same spectrum.
	
	Deciding between these prescriptions, and assessing the effectiveness of NLLD overall, requires first-principles kinetic simulations that simultaneously resolve the CR gyroscale and the thermal plasma physics responsible for Landau damping. Here the literature remains sparse, offering little guidance.
	The computationally efficient MHD-PIC framework \citep{bai2019} treats the thermal plasma as an ideal MHD fluid and thereby precludes Landau damping, and with it NLLD, by construction.
	Full PIC simulations \citep{holcombMicrophysicsGyroresonantStreaming2019,shalabyNewCosmicRaydriven2020,shalaby2026} include NLLD self-consistently, but need to make stronger compromises in the simulations parameters. So far, they did not find that it saturated the instability. Although the investigation of NLLD by \citet{shalaby2026} was restricted to an unrealistic parameter regime, they validated the fluid-PIC method against fully kinetic PIC simulations. In the present work, we use the fluid-PIC method to extend their findings into a more physically motivated parameter regime.
	In fluid-PIC simulations with Landau closures, \citet{lemmerz2024CouplingMultifluidDynamics} also found that the instability saturated not through NLLD, but through gyrophase bunching \citep{lemmerz2025}, although using a simplified gyrotropic ring distribution for CR.
	Recent hybrid-PIC simulations \citep{schroer2025,schroer2026} instead affirm the role of NLLD as a saturation mechanism using a monoenergetic shell distribution, while also finding that it drives a strong inverse cascade that should theoretically be absent if only CR streaming modes are present. 
	
	Because the investigated damping process is nonlinear, it often remains ambiguous which features of a simulation result from NLLD, from another physical process, or from numerical artifacts.
	We remove this ambiguity by making direct side-by-side comparisons to evaluate the effectiveness and influence of NLLD on the streaming instability: using the fluid-PIC method, in which the ambient electrons and ions are evolved as fluids with Landau closures, we switch Landau damping -- and with it NLLD -- on and off while holding everything else fixed.
	Because the strength of NLLD depends critically on the intensity of the interacting Alfv\'en modes, we perform this comparison both for simulations in which all wave power is generated by the CRs themselves, and for simulations in which large-scale modes are seeded by hand to mimic external turbulence.
	
	The paper is structured as follows.
	Section~\ref{sec:theory} gives a brief and pedagogical overview of the theory of NLLD, tailored specifically to CR streaming. Section~\ref{sec:method} explains the numerical method and the simulation setup, and Section~\ref{sec:results} presents an in-depth comparison of the simulations and their results.
	We discuss the limitations of this work and compare it with previous literature in Section~\ref{sec:discussion}, before summarizing our conclusions in Section~\ref{sec:conclusion}.

	\section{Theory}
	\label{sec:theory}

	This section provides the minimal theoretical background needed to interpret our simulations.
	Section~\ref{sec:theory_notation} introduces the wave-mode designations and Els\"asser variables that underlie all spectral diagnostics of this paper.
	Section~\ref{sec:theory_streaming} applies NLLD to the CR streaming instability and derives the balance between wave growth and NLLD.
	More detailed derivations and extended discussions are deferred to a companion theory paper (Lemmerz in prep.).

	\subsection{Wave-mode Designation: Propagation, Rotation, and Els\"asser Variables}
	\label{sec:theory_notation}
	Throughout this paper, both the wavenumber $k$ and the frequency $\omega$ of a mode are signed quantities.
	A parallel-propagating Alfv\'en wave then carries two independent binary labels.
	The first is its \emph{propagation direction} $\sign(\omega/k)$: the wave travels parallel (at wave velocity $\vwa = +\valf$, ``forward'') or antiparallel ($\vwa = -\valf$, ``backward'') to the background field $\bm{B}_0$, which we take to point along $+\hat{\bm e}_\parallel$ so that $\valf > 0$.
	We reserve the symbol $\pm$ for labeling the propagation direction, matching its use in the Els\"asser variables $z^{\pm} = \varv_{i,\perp} \pm B_\perp/\sqrt{\mu_0 \rho}$, where $\mu_0$ is the vacuum permeability and $\rho$ is the plasma mass density.\footnote{The Els\"asser superscript is opposite in sign to $\sign(\omega/k)$: the linearized Els\"asser equations for an inviscid and isobaric fluid $\partial_t z^{\pm} \mp (\bm \varv_\mathrm{A}\bm\cdot\bm\nabla)\, z^{\pm} = 0$ show that a $z^{\pm}$ mode has phase velocity $\vwa = \mp\valf$, i.e., $z^{+}$ denotes backward- and $z^{-}$ forward-propagating waves.} Decomposing the wave spectrum into Els\"asser variables thus allows us to determine the energy carried by forward- and backward-propagating Alfv\'en waves in our simulations.
    While they are conventionally full 3D vectors, throughout this paper it is sufficient to analyze the 2D plane perpendicular to the magnetic field, making it more convenient to define them as complex quantities, shown next.
    
	The second label is the \emph{helicity} $\sign(k)$, i.e., the sense in which the perturbation vector rotate along $\bm{B}_0$ at a fixed time. We differentiate the sign of $k$ in the simulations by Fourier transforming the complex variable $\varv_\perp = \varv_y - \ci \varv_z$, and $B_\perp$ is defined analogously, which both enter into our definition of the complex $z^\pm$.  
	Finally, the sign of the rotation frequency $\omega$ follows from the dispersion $\omega = k\vwa =  k \left(\mp \abs{\valf}\right)$.\footnote{\citet{holcombMicrophysicsGyroresonantStreaming2019} refers to $\mp \abs{\valf}$ as anti-parallel ($-$) and parallel ($+$), and designate the sign of $\omega$ as left-handed (ion-rotation) and right-handed (electron-sense). Their notation uses $\varv_\perp = \varv_y \textbf{+} \ci \varv_z$, corresponding to a mapping of $(\omega, k)_\mathrm{Holcomb}\to (-\omega, -k)$ here. Another common convention \citep{stix1992} reserves $\pm$ for the polarization, using the basis $B_\perp^\pm=B_y \pm \ci B_z = \abs{B_\perp}\exp(\pm \ci (k x-\omega t))$, and typically limiting either $k>0$ or $\omega>0$. This can be mapped to our convention where both $k$ and $\omega$ are signed quantities, see ~\citet{weidl2019BackgroundgyroresonantCharacterBells}.}
	A linearly polarized wave is the superposition of two modes of equal amplitude and phase velocity, but opposite $k$. 

	\textbf{The sign of $k$} (helicity) is important for the resonance condition.
	A CR with velocity parallel to the background magnetic field, $\varv_\parallel = \mu \varv$, where $\mu$ is the pitch-angle cosine, and relativistic gyrofrequency $\Omega_\crrm = \Omega_i/\gamma_\crrm$, where $\gamma_\crrm$ is the Lorentz factor and $\Omega_i$ is the ion gyrofrequency, is gyroresonant with a mode if
	\begin{equation}
		\varv_\parallel - \vwa = \frac{\Omega_\crrm}{k},
		\label{eq:res_theory}
	\end{equation}
	so that energetic CRs much faster than the wave velocity, $\varv \gg \abs{\vwa} = \valf$, approximately fulfill
		\begin{equation}
		\varv_\parallel= \mu \varv \approx \frac{\Omega_\crrm}{k},
		\label{eq:res_approx}
	\end{equation}
	showing that forward-moving CRs ($\mu>0$) resonate with $k>0$ modes and backward-moving CRs ($\mu<0$) with $k<0$ modes. While the gyroresonance condition is formulated for individual particles, it is straightforward to generalize it to distributions. For example, a CR distribution that is isotropic in some drifting frame spans $\mu\in[-1,1]$, where the forward-moving part, $\mu\in(0,1]$, of the distribution resonates with a spectrum of $k>0$ modes, while the backward-moving part $\mu\in[-1,0)$ resonates with a spectrum of $k<0$ modes.  \\
	\textbf{The sign of $\omega$} sets the sense of rotation with $\omega>0$ indicating right-handed (electron -like) rotation and $\omega<0$ indicating left-handed (ion -like) rotation. The sign is also important for NLLD, as should become evident in Section~\ref{sec:theory_LV}. It determines whether a wave is damped or growing due to nonlinear Landau interactions.\\
	\textbf{The sign of $\vwa=\omega/k$}, or the propagation direction of the wave, is also important for the strength of NLLD, but even more so for pitch-angle scattering of the CRs. Positive $\valf$ will scatter CRs in a frame that is different from waves at $-\valf$, ultimately deciding on whether CRs isotropize with either $\vdr\to+\valf$ or $\vdr\to-\valf$, or in between (see Section~\ref{sec:drift}).

\begin{figure}
    \centering
    \includegraphics[width=1\linewidth]{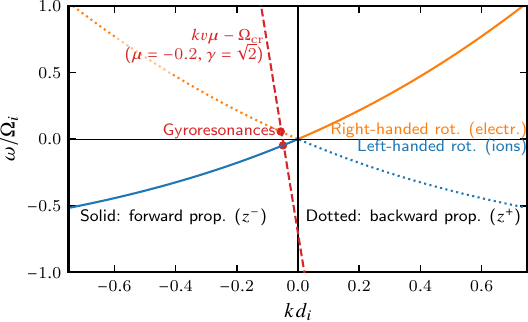}
    \caption{	Dispersion relation of transverse waves propagating along $\bm{B}_0$ in the fluid model. Toward large scales, $k d_i\ll1$, they fulfill the Alfv\'en wave dispersion relation $\omega = \mp k \valf$.
	Orange curves are modes with $\omega>0$ (right-handed, electron-sense rotation); blue curves have $\omega<0$ (left-handed, ion-sense rotation).
	Solid (dotted) lines denote forward (backward) propagation in the sense $\sign(\omega/k)=+ (-)$, i.e.\, these are $z^-$ ($z^+$) Els\"asser modes. The red dots mark the intersections of the red dashed line, $\omega = k\varv\mu - \Omega_{\mathrm{cr}}$, with the dispersion
relation $\omega(k)$ and thus satisfy the
gyroresonance condition~\eqref{eq:res_theory} for an example
particle. Particles with negative pitch-angle cosine $\mu$ resonate
with waves with negative wavenumber $k$. Our CR distribution spans
$\mu \in [-1,1]$ and therefore excites a spectrum of waves.
 For this visualization, we adopted the parameters $m_i/m_e=25$ as used in our simulations.}
    \label{fig:convention}
\end{figure}

    Figure~\ref{fig:convention} provides a visualization of the dispersion relation and gyroresonance condition in this convention.

	\subsection{NLLD Applied to CR Streaming: Growth-Damping Balance}
	\label{sec:theory_streaming}
	In equilibrium, growth by the streaming instability is balanced by damping, or
	\begin{equation}
		\Gamma_\mathrm{str} (k) + \Gamma_\mathrm{NLLD}(k)  = 0.
		\label{eq:equilibrium_toplevel}
	\end{equation}
	In the following we give the instability growth rate 	$\Gamma_\mathrm{str}$ and the NLLD rate $\Gamma_\mathrm{NLLD}$ before reformulating Eq.~\eqref{eq:equilibrium_toplevel} in a way that allows us to compare with the simulated wave spectra.

	\subsubsection{Streaming Instability Growth Rate}
	CRs drifting at $\vdr > \valf$ along $\bm B_0$ render Alfv\'en waves unstable \citep{wentzel1968HydromagneticWavesExcited,kulsrud1969}. For a monoenergetic shell distribution $f=n_\crrm \delta(u-u_0)/(4\pi u^2)$ of CRs that is isotropic in the reference frame drifting with $\vdr$, which we use in our simulations, a derivation following \citet{zweibel2003} (Appendix~\ref{app:shell_growth}) yields for the streaming instability growth rate
	\begin{equation}
		\Gamma_\mathrm{str}(k) =  \frac{\pi}{4}\, \frac{n_\crrm}{n_i}\, \abs{\frac{\Omega_i}{k}}\, \frac{1}{u_0} \left(\frac{ \vdr}{\mp\valf} - 1\right) \Theta\!\left(u_0 - \abs{\frac{\Omega_i}{k}}\right) \Omega_i,
		\label{eq:growth_shell}
	\end{equation}
    where $n_s$ is the number density of ions ($i$) or CRs ($\crrm$), the proper velocity is $u=\gamma_\crrm \varv$, which is set by the parameter $u_0$, and $\Theta$ is the Heaviside step function.
    
	We highlight three points that are necessary to understand the instability evolution below:
	\begin{enumerate}
	    \item 
 only resonant \emph{forward-propagating} $z^-$ waves grow once $\vdr>\valf$, while resonant \emph{backward-propagating} $z^+$ waves are damped by the CRs.
	\item The growth is symmetric in helicity, $\Gamma_\mathrm{str}(k) = \Gamma_\mathrm{str}(-k) \propto 1/\abs{k}$, so streaming drives \emph{linearly polarized} waves\footnote{Because the circularly polarized modes $B_\perp(k)$ and $B_\perp(-k)$ are excited with the same growth rate, their superposition appears as a linearly polarized wave, $B_\perp(x) = B_\perp(k) \exp(\ci k x) + B_\perp(-k) \exp(-\ci k x)
\approx  B_\perp(k) \left[\exp(\ci k x) + \exp(-\ci k x)\right] = 2 B_\perp(k) \cos(k x) \overset{!}=B_y(x) + \ci B_z(x)$ implying oscillation in $B_y$ and vanishing $B_z=0$, up to an arbitrary angle in the $y$-$z$ plane.};
    this is a consequence of the symmetry between forward and backward moving CRs that excite $+k$ and $-k$ modes, respectively, at equal strength.
	\item Only scales below the maximum CR gyroradius are unstable, $\abs{k} \geq \kmin = \abs{\Omega_i}/u_0$, so the self-generated spectrum contains no power at larger scales.
    	\end{enumerate}
    
    All but the last point (3) are also true for a CR power-law distribution, as long as all other approximations are fulfilled as well.
	\subsubsection{NLLD of Two Interacting Alfv\'en Waves}
		\label{sec:theory_LV}
	Two Alfv\'en waves $(\omega_1, k_1)$ and $(\omega_2, k_2)$ beat and exert a longitudinal ponderomotive force on the plasma, driving a compressive beat wave that travels at $\vbeat = \Delta\omega/\Delta k$, where $\Delta\omega = \omega_1 - \omega_2$ and $\Delta k = k_1 - k_2$ \citep{hollweg1971DensityFluctuationsDriven,hollweg1971NonlinearLandauDamping}.
	Thermal particles moving at velocities somewhat smaller than $\vbeat$ Landau damp this beat wave, draining energy from the parent Alfv\'en waves in the process called NLLD. 

	\citet{lee1973DampingNonlinearWaveparticle} (hereafter LV) derived the resulting damping rate for two interacting Alfv\'en wave modes $\Gamma_1$ and $\Gamma_2$, and found that NLLD conserves the \emph{wave action} of both waves, \begin{equation}
		\Gamma_1 B_1^2/\omega_1 + \Gamma_2 B_2^2/\omega_2 = 0,
		\label{eq:wave_action_conservation}
	\end{equation}
where subscripts $1$ and $2$ denote that it is a quantity of wave $1$ or $2$, respectively.
    
    The wave action, $B_\perp^2/\omega$, may be interpreted as an angular pseudomomentum \citep{andrews1978,bliokh2022}. Yet, perhaps counterintuitively, a wave rotating faster than another carries less of that angular pseudomomentum at the same intensity. 
	Because this angular pseudomomentum is conserved overall, there are two possible but fundamentally different interactions, one corresponding to colliding wave packets and the other to co-propagating wave superposition.
    The first is an inelastic collision of two oppositely rotating wave packets ($\sign\omega_1 = -\sign\omega_2$): their signed pseudomomenta can (partially) cancel, allowing both waves to damp while their energy is thermalized, meaning that $\Gamma<0$ for both. The second involves co-rotating wave packets ($\sign\omega_1 = \sign\omega_2$), whose pseudomomenta have the same sign and therefore add up rather than cancel. In this case, to conserve the total wave action, one of them has to grow in intensity instead of damp.
	Consequently, a faster-rotating wave $2$ will transfer a fraction $\omega_1/\omega_2$ of its lost energy to the slower-rotating -- and hence larger-scale -- Alfv\'en wave $1$, while the remaining energy heats the background plasma. NLLD acts as a dissipative inverse cascade in this case, transferring energy from smaller scales to larger scales.
	Thus, despite being called nonlinear Landau \emph{damping}, not all modes are damped, but some grow.
	
	We omit the details of the damping rates for brevity, and focus on the case of a magnetic spectrum generated by the CR streaming instability in an isothermal plasma.
	\subsubsection{NLLD of a Streaming Instability Generated Spectrum in an Isothermal Plasma}
    As shown above, the streaming instability generates a linearly polarized spectrum of co-propagating waves only, and we limit ourselves to this case.
	A straightforward way to extend the two-wave LV interactions to a spectrum is to integrate them over the wave spectrum of all possible interaction partners of a mode $k$. 
	
	First, however, we look at the case of a three wave interaction of co-propagating waves that is instructive for the streaming instability. We look at a larger-scale Alfv\'en mode oscillating at $\omega_1$ interacting with a linearly polarized Alfv\'en wave oscillating at $\omega_2$ ($\omega_1<\omega_2$), which in reality consists of two circularly polarized modes $\omega_2$ and $\omega_{-2}=-\omega_2$ with equal intensities $B_2^2=B_{-2}^2$.
	The total NLLD rate of wave $1$ is the sum of the individual interactions with wave $2$ and $-2$, so $\Gamma_1 = \Gamma_{1,\text{due to }2}+\Gamma_{1,\text{due to }-2}$. 
	Leaning on the previous arguments around Eq.~\eqref{eq:wave_action_conservation}, we find that wave 1 is damped and fed by the other waves at equal rates, $\Gamma_{1,\text{due to }2}=-\Gamma_{1,\text{due to } -2}$, and the net growth/damping of wave 1 vanishes, $\Gamma_1=0$.
	Thus, the faster-rotating linearly polarized wave has no net impact on wave $1$, and no inverse cascade is possible.
	Nevertheless, waves $2$ and $-2$ are both damped due to interaction with wave $1$ (and also due to interaction with each other). Consequently, damping is provided only by wave power at larger scales in this configuration, while larger-scale waves are effectively \emph{un}damped by smaller scale waves.
	
	\citet{miller1991} formalized this for wave spectra of linearly polarized wave modes, giving the damping rate as
	\begin{align}
		\Gamma_\NLLD^{\mathrm{co}}(k) &\simeq -\frac{ F(\beta_i, \beta_e)}{8}\, \abs{k}\, \valf\, I^{\mathrm{co}}_{<k}\label{eq:NLLD_iso} , \\
		I^{\mathrm{co}}_{<k} &= \int_{\abs{k'} < \abs{k}} \frac{\epsilon^{\mathrm{co}}(k')}{B_0^2}\, \mathrm{d}k',
	\end{align}
	where $\epsilon^\mathrm{co}$ is the wave spectral density of only the co-propagating Alfv\'en waves and $F(\beta_i, \beta_e)$ \citep[Equation 3.2 of ][]{miller1991} is a factor that sets the NLLD strength depending on the ratio of the beat wave velocity and the thermal velocities $\varv_{\mathrm{th},s}$ of the ions ($s=i$) and electrons ($s=e$). Because the beat wave of co-propogating waves travels at the Alfv\'en speed, this ratio is fully encapsulated by the plasma $\beta_s=2\varv_{\mathrm{th},s}^2/\valf^2$ of the ions and electrons.
	The net effect of NLLD in this case is a pure dissipation of the smaller-scale waves by larger-scale waves, without any reverse effect. 
	Somewhat surprisingly, there is no locality weighting in $k$; a wave is damped by every larger-scale wave, at a rate that depends only on that wave's intensity. All large-scale waves enter the integrated intensity $I_{<k}^\mathrm{co}$ equally. 

	\subsubsection{Streaming Instability Growth--Damping Balance}
    The streaming instability excites only linearly polarized forward-propagating waves, so all waves are co-propagating. Thus, we may use Eq.~\eqref{eq:NLLD_iso} to estimate an equilibrium.
	The growth rate~\eqref{eq:growth_shell} decreases strictly monotonically with $\abs{k}$, while the damping \eqref{eq:NLLD_iso} increases strictly monotonically with $\abs{k}\, I^{-}_{<k}$; therefore, the two can balance at most at a single wavenumber $k_\mathrm{bal}$.
	A marginal state in which growth and damping cancel over a \emph{range} of $k$ would require a monotonically increasing growth rate $\Gamma_\mathrm{str}(k)$, which can be fulfilled by some power-law CR distributions. In this paper, however, we focus on CRs with a shell distribution only.
	NLLD should thus truncate the spectrum: modes with $\abs{k} > k_\mathrm{bal}$ are damped, modes with $\abs{k} < k_\mathrm{bal}$ continue to grow, and equating \eqref{eq:growth_shell} with \eqref{eq:NLLD_iso} yields the cutoff scale $k_\mathrm{bal}$ where damping and growth balance
	\begin{align}
		\Gamma^\mathrm{co}_\NLLD (k_\mathrm{bal}) = \Gamma_\mathrm{str}(k_\mathrm{bal})\\
\Leftrightarrow		\left(\frac{k_\mathrm{bal}}{\kmin}\right)^{2} = 2\pi \frac{ \dfrac{n_\crrm}{n_i}\, \dfrac{u_0}{\valf}\left(\dfrac{\vdr}{\valf} - 1\right)}{ F(\beta_i, \beta_e)\, I^{\mathrm{co}}_{<k_\mathrm{bal}}}.
		\label{eq:kbal}
	\end{align}
	
	As all modes with $k<k_\mathrm{bal}$ remain growing, $I^{\mathrm{co}}_{<k_\mathrm{bal}}$ keeps growing as well. Thus, if NLLD is the main saturation mechanism, we also expect the cutoff $k_\mathrm{bal}$ to decrease slowly over time until it becomes the largest resonant mode, $k_\mathrm{bal} \rightarrow k_{\mathrm{min}}$, and all smaller-scale modes are damped. 
	
	In contrast, the quasilinear scattering rate is given by \citep{kulsrud1969}
    \begin{equation}
	    \nu_\mathrm{QLT}(k) = \pi\Omega \abs{k} \frac{\abs{\epsilon}(k)}{B_0^2},
        \label{eq:nu_qlt}
	\end{equation} 
    where the scattering rate of a resonant CR depends only on the spectral wave energy density $\epsilon(k)$ at the scale matching its resonant $k$, where $\int \epsilon(k) \mathrm{d}k= \abs{B_\perp}^2$. If wave growth is instead interrupted by quasilinear pitch-angle scattering, the saturated spectrum follows from
	\begin{equation}
		\nu_\mathrm{QLT}(k) \approx \Gamma_\mathrm{str}(k) \quad\Rightarrow\quad \abs{B_\perp}(k) \propto \abs{k}^{-1},
        \label{eq:qlt_spectrum}
	\end{equation}
	This $1/\abs{k}$ shape is indistinguishable from that of the growth rate itself (Eq.~\ref{eq:growth_shell}), but clearly different from the spectrum expected for NLLD, which gives us one observable to distinguish the two.
	As another indicator of the effectiveness of NLLD, our fluid-PIC method (Section~\ref{sec:method}) allows toggling Landau damping on and off, thereby switching NLLD off entirely and allowing for a direct side-by-side comparison.
	
	\subsubsection{NLLD of Waves with Mixed-propagation Directions}
	NLLD between counter-propagating waves is more complicated than that of co-propagating waves, because the beat wave velocity is no longer just the Alfv\'en speed, but depends on the relative $k$ vectors of both waves.
	A forward mode $k_1$ and a backward partner $k_2$ ($\omega_2 = -k_2\valf$) beat at $\vbeat = \valf\, (k_1+k_2)/(k_1-k_2)$, so the plasma response must be evaluated pair by pair and the expressions quickly become unwieldy. We therefore do not reproduce the spectral damping rate for the more general case here. Instead, we want to emphasize two points that are relevant for our results: First, the NLLD rate of counter-propagating waves can be more faster than that of co-propagating waves and second, the inverse cascade is not subject to the above cancellation argument presupposing a linearly polarized co-propagating spectrum.
    
The second point follows quite intuitively when looking at a backward propagating mode that interacts with a linearly polarized forward propagating mode.
As in the co-propagating case, one of the two interactions is co-rotating and the other counter-rotating, so their contributions to $\Gamma(k)$ enter with opposite signs. Now, however, they are weighted by the plasma response at two \emph{different} beat velocities and therefore no longer cancel.

The first point, that mixed propagation can lead to more efficient NLLD, cannot be proven without deeper analysis of the LV formula that we omit here, but at least can be made plausible. For mixed propagation, the beat wave velocity is no longer locked to the Alfv\'en speed. Instead, a broad spectrum of forward propagating waves interacting with one backward moving waves will excite a broad spectrum of beat waves with different velocities.
While some interacting pairs may experience little NLLD, others, specifically those that generate beat waves near the ion's thermal velocity $\abs{\vbeat} \approx \varv_{\mathrm{th},i}/\sqrt{2}$, will experience a large NLLD rate, because this is where the ion Landau damping response peaks. Beat waves excited at the electron thermal velocity are efficiently damped as well, which is more likely in our simulations because we adopt a reduced mass ratio of $m_i=25 m_e$, which implies an artificially small scale separation between the electron and ion thermal velocities.
	
	\section{Method}
	\label{sec:method}
	\subsection{Fluid-PIC Method and NLLD}
		\begin{table*}[t]
	\centering
	\caption{Overview of simulations. Shared parameters: $\beta_i=\beta_e=1$, $m_i/m_e=25$, $n_\crrm/n_\mathrm{bg}=5\times10^{-4}$, $\valf=10^{-2}c$. The CRs are initialized as a monoenergetic shell distribution of $f_\crrm = n_\mathrm{cr} \delta(u-u_0)/(4\pi u^2)$ where the proper velocity is $u_0=c$ (corresponding to $\gamma_\crrm=\sqrt{2}$ and thus ordinary velocity of $\varv_0\approx0.7c=70\valf$) drifting at a speed of $\vdr=5\valf$ relative to the background frame. The grid resolution is $\Delta x=0.45c/\omega_p\approx 0.088 d_i$, where $\omega_p$ is the plasma frequency, and the timestep is $\Delta t=0.45\Delta x/c$.
	}
	\label{tab:simulations}
	\begin{tabular}{l|cccccc}		
		Label & Landau closure & Seeded modes & CR particles per $d_i$ & $L_x [d_i]$ & $L_x[ 2\pi /\kmin]$ & Grid Cells \\[0.1cm]
		\hline
		NoLD & off & none & 4080 & 20,744.84 & ~33 & 235,063\\
		LD & on & none & 4080 & 20,744.84 & ~33 & 235,063\\
		NoLD low-ppc & off & none & ~~204  & 62,863.14 & 100 & 712,312\\
		LD low-ppc & on & none & ~~204 & 62,863.14 & 100 & 712,312\\
        NoLD+ext & off & $n_w=20$ & 4080 & 20,744.84 & ~33 & 235,063\\
		LD+ext & on & $n_w=20$ & 4080 & 20,744.84 & ~33 & 235,063\\
	\end{tabular}
\end{table*}
	We use the fluid-PIC method as described in \citet{lemmerz2024CouplingMultifluidDynamics} implemented as an extension of the SHARP PIC code \citep{shalaby2017,shalabyNewCosmicRaydriven2020}. It models the ambient electrons and ions using a full evolution of the fluid equations for each species separately, recovering the kinetic effect of Landau damping through Landau closures. CRs are treated fully kinetically with the PIC method, and the coupling between the individual species is achieved solely through Maxwell's equations. We use the Landau closure $R_{3,1}$ as given by \citet{hunana2018NewClosuresMore} for both fluid species (ions and electrons) and perform a full spectral analysis at each timestep to accurately capture the Landau damping rate at all resolved $k$.
	Our NoLD runs are performed without the Landau closure, using a non-dissipative ideal gas closure instead. As a result, Landau damping and NLLD are absent in these simulations.

	While Landau fluids (LFs) recover Landau damping well in the linear regime, they lack some kinetic effects that are important to keep in mind. 
	Landau damping leads to a flattening of the distribution function due to particle trapping, which potentially reduces the (non-linear) Landau damping rate \citep{volkCesarsky1982} over time. It has been argued, however, that this trapping is likely unstable and the unsaturated damping rate applies \citep{lagageCesarsky1983, miller1991,schroer2026}, which LFs reproduce accurately.
	
	Furthermore, even though linear Landau damping is captured by LFs, it is a priori unclear whether NLLD is captured accurately as well \citep{mattor1992CanLandaufluidModels}.
	In LF models, NLLD can be understood by likening the electrostatic ion-acoustic wave to a damped oscillator that is forced by the ponderomotive force of two interacting Alfv\'en waves. NLLD is therefore most efficient when the beat wave resonates with the ion-acoustic wave,  $\varv_\mathrm{beat}\sim \varv_\mathrm{IA}$, which is when the plasma $\beta\sim 1$ (assuming copragating waves), and its efficiency drops as $\beta\gg 1$. 
	
	On the other hand, kinetic models become more deeply resonant in the $\beta>1$ regime and predict that NLLD scales roughly with $\sqrt{\beta}$, because $F(\beta_i,\beta_e)$ entering Eq.~\eqref{eq:NLLD_iso} scales with $F(\beta_i,\beta_e)\approx\sqrt{\beta_i}\exp(-1/\beta_i)\sim\sqrt{\beta_i}$ to leading order in $\beta_i$ in the high $\beta_i$ limit. 
	This is because kinetic models take the ponderomotive force on each particle into account, including individual particle resonances instead of wave-wave resonances and allowing for corrections to the Larmor radius that are all lacking in our LF model, which simply analyzes the wave spectra.

	Thus, NLLD in LF and kinetic models differ in the regime of $\beta\gg1$, where LFs underestimate the damping by a factor of roughly $\beta$ \citep{mattor1992CanLandaufluidModels}. 
	We are interested in the regime around $\beta=1$, where LF models reproduce the kinetic behavior of NLLD accurately, motivating the use of the fluid-PIC code. 
	We choose $\beta_i=\beta_e=1$ in the simulations presented in this paper.

\subsection{Simulation Setup}

	We perform 1D periodic-box simulations that retain all three velocity components. A uniform background magnetic field is initialized along the box's $x$ direction, with a magnitude corresponding to $\valf = 10^{-2}c$. This setup represents CRs streaming along a magnetic field line: they are initially unstable because of their drift and are then allowed to relax. This setup thus models a cloud of CRs propagating along the magnetic field, as happens, for example, when CRs stream from a galactic disk into the halo and drive a galactic wind. The CR distribution is a monoenergetic distribution that is isotropic in a drifting frame. Our parameters for the different simulations are given in Table~\ref{tab:simulations}. We do not simulate any CR electrons; the initial parallel CR current is neutralized by initializing the background electron fluid with a drift of $n_\mathrm{cr}\left<\varv_x\right>/n_\mathrm{bg} \approx 2\times 10^{-5}c$. In earlier simulations featuring comparable CR currents we found no discernible difference between boosting background electrons or using an electron beam to neutralize the drifting ions \citep[Figure 14 of][]{lemmerz2025}. 
    Our simulation frame is comoving with the background ions.

  	     \begin{figure*}[t!]
    \includegraphics[width=\textwidth]{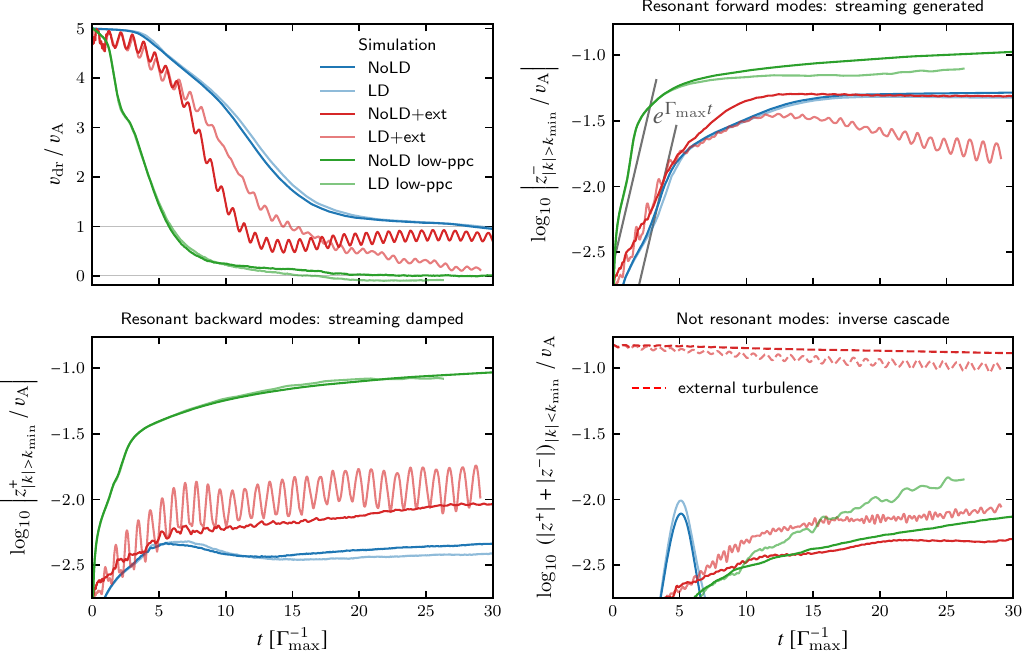}
    \caption{Comparison of the evolution of the macro quantities of each simulation. We show (top left) the CR drift velocity, which tends to either the Alfv\'en velocity $\valf$ or the background velocity $0$, (top right) gyroresonant $z^-$ which measures the intensity of forward-moving, streaming excited Alfv\'en waves, (bottom left) gyroresonant $z^+$ which measures the evolution of backward propagating waves, which the streaming instability cannot excite, (bottom right) intensity of Alfv\'en waves that are at scales larger than the maximal gyroradius and therefore not resonant with any CRs.
    For this figure, we use the shorthand notation 
    $\abs{z}^\pm_{\abs{k}>\kmin} = \left(\sum_{k=\kmin}^{\infty} + \sum_{k=-\kmin}^{-\infty}\right)  \abs{z^\pm_k}$. In the +ext simulations only, dashed lines show the amplitudes of the manually seeded waves separately (external turbulence) and $\abs{z}^\pm_{\abs{k}>\kmin}$ excludes $k_\mathrm{ext}$.
}
    \label{fig:zpanels}
\end{figure*}
    
	The two baseline simulations \emph{NoLD} and \emph{LD} are performed to judge NLLD due to self-generated CR modes only.
	We do not initialize any fields, with the exception of the \emph{+ext} simulations, where we initialize the Alfv\'en wave modes $z^+(k)$, $z^-(k)$, $z^+(-k)$, $z^-(-k)$ with equal amplitudes of $\delta B_\perp=7.5\% B_0$ at box harmonic $n=20$, whereas the largest resonant scale corresponds to $n=33$. To initialize the waves in the simulation, we solve the dispersion relation for an electron-ion-fluid as an eigenvalue problem \citep{xie2014PDRFGeneralDispersion}. We then initialize the simulation from the resulting eigenvectors, which are made up of the electromagnetic field components $\bm B, \bm E$ and the electron and ion fluid quantities (density, bulk velocities, and pressure). 
    
	In our low-ppc simulations we reduce the number of computational particles per cell and extend the box size, although we have repeated the low-ppc simulations at smaller box sizes and find no notable differences. 
	
	While we run most simulations until, or even after $t=30 \Gamma_\mathrm{max}^{-1}$, some are run for a shorter time. These shorter runs are LD simulations with strong co- and counter-propagating modes, which generate rotational discontinuities that are steady-state solution of the LF system with these modes \citep{medvedev1996,medvedev1997}. These discontinuities require strong transverse current gradients, which are supported by growing electron and ion density spikes (with no net charge separation).
	Instead of choosing smaller and smaller timesteps to allow for these rapidly growing density imbalances that arise from the wave-wave interactions, we end our simulations early and focus on the CR interactions up to this point.

    \subsection{Simulation Parameter Ordering}
    \label{sec:sim_parameter_ordering}
    Our simulations are set up to reproduce key timescale orderings of the realistic parameters in the ISM.
 Although the Alfv\'en velocity of $10^{-2}c$ is artificially high, the largest gyroresonant scale $\kmin d_i = \valf/u_0 = 0.01 \ll 1$ is firmly in the MHD limit where Alfv\'en waves can be excited, where $d_i$ is the ion skin depth. 
        
     The simulation parameters also ensure that all waves' growth rates are slower than their rotation rates, as even the maximum growth rate is an order of magnitude smaller than its rotation rate \begin{equation}\Gamma(k_\mathrm{max})\approx1.5\times10^{-3} \Omega_i \ll \omega(k_\mathrm{max}) = 10^{-2} \Omega_i.
     \end{equation}
    
    We also estimate the competing NLLD rate for the LD+ext simulation from Eq.~\eqref{eq:NLLD_iso}, where we only take our externally seeded co-propagating wave energy into account. The counter-propagating waves provide additional damping, but for simplicity we neglect them in the following. With an order unity $F(\beta_i, \beta_e) = 1.88$, and the spectra $\epsilon^\mathrm{co} (k) = (0.075 B_0)^2 \delta(\abs{k}-k_\mathrm{ext})$, we find that $\Gamma^{\mathrm{LD+ext}}_\mathrm{NLLD}(\kmin) \approx 2.6\times10^{-5}\Omega_i$, so roughly two orders of magnitude less than the streaming growth rate for $\kmin$ at initialization. 
    Thus, NLLD could theoretically balance growth in two ways. Either, the drift velocity reaches $\varv_\mathrm{dr}^\mathrm{LD+ext}\approx 1.07\valf$, which is when the maximum growth rate (Eq.~\ref{eq:growth_shell}) matches the level of $\Gamma^{\mathrm{LD+ext}}_\mathrm{NLLD}(\kmin)$ and all growth should stop; however, this is not what we observe in Section~\ref{sec:res_ext}. Alternatively, NLLD curbs growth at smaller scales, as predicted by Eq.~\eqref{eq:kbal}. This should happen at 
    \begin{equation}
    \label{eq:kbal_ldext}
    k_\mathrm{bal}^\mathrm{LD+ext} \approx 7.7 \kmin
    \end{equation}
    from the beginning of the simulation.

	\section{Results}
	\label{sec:results}

	All diagnostics in this section build on the wave-mode designations of Section~\ref{sec:theory_notation}.
	The Els\"asser fields separate the transverse fluctuation energy by propagation direction: $z^-$ measures forward modes, which co-propagate with the drifting CRs and are the only resonant modes that streaming amplifies (Eq.~\ref{eq:growth_shell}), whereas $z^+$ measures backward modes, which resonant CRs damp.
	The streaming instability only generates resonant $z^-$ (forward moving) Alfv\'en modes; any other modes therefore indicate other generation mechanisms.

	Figure~\ref{fig:zpanels} compares the macroscopic evolution of the six runs of Table~\ref{tab:simulations} and contains the central result of this paper.
	In all runs the streaming-generated wave amplitude grows initially at roughly the maximum rate $\Gamma_\mathrm{max}$ predicted by Eq.~\eqref{eq:growth_shell} (top right panel), as long as the average amplitude is dominated by the fastest growing waves excited at the scale $\kmin$. However, the low-ppc simulation grows slightly faster while the +ext and baseline runs (NoLD, LD) grow slightly slower.
	The subsequent evolution separates the runs into two families, but the split is not set by whether the Landau closure is turned on or off.
	In the baseline runs and in NoLD+ext, backward wave power stays more than an order of magnitude below the forward power, and the CR drift relaxes to the Alfv\'en speed: the CRs isotropize in the frame of the forward-propagating Alfv\'en waves they generated.
	In LD+ext and in both low-ppc runs, by contrast, resonant backward modes match the forward modes in power, and the drift velocity drops below $\valf$ and approaches the background rest frame, $\vdr\to0$.
    
	NLLD only has a major impact in the +ext simulations that manually seed large intensity Alfv\'en modes at scales not resonant with the CRs, while minor differences can be seen in the wave intensities of the low-ppc simulations, and virtually no differences are found in the baseline simulations. This clearly shows, that NLLD is too weak to curb the streaming instability growth in the absence of large-scale turbulence in our simulations.
	
	We examine the three cases in turn: self-confinement in isolation (Section~\ref{sec:res_self}), spurious modes caused by low macroparticle resolution (Section~\ref{sec:res_lowppc}), and externally imposed large-scale modes (Section~\ref{sec:res_ext}).

		 	\begin{figure*}[t]
		\includegraphics[width=\textwidth]{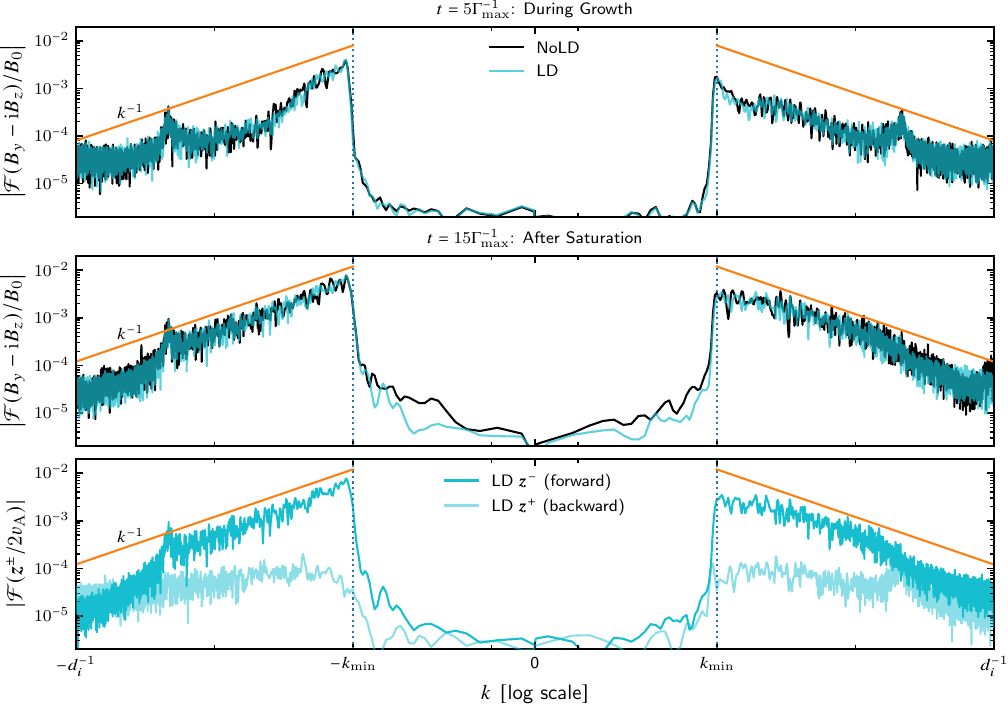}
		\caption{Transverse magnetic spectra of the baseline runs during growth (top) and after saturation (middle). LD and NoLD are indistinguishable at both times; the spectrum follows $1/\abs{k}$ (orange) over the resonant range $\kmin \leq \abs{k} \lesssim d_i^{-1}$ and virtually no power develops at larger scales, which are not resonant. The bottom panel shows a decomposition of the LD spectrum into Els\"asser variables, showing that almost exclusively forward moving waves are excited, as expected by the linear dispersion relation. Toward $\abs{k}\lessapprox d_i^{-1}$ the decomposition becomes less reliable, as a consequence of the wave dispersion changing.}
		\label{fig:spectra_baseline}
	\end{figure*}
    
	\subsection{NLLD due to Self-generated Alfv\'en Waves Only}
	\label{sec:res_self}

	Figure~\ref{fig:spectra_baseline} shows the spectra of the Alfv\'en wave during growth and after saturation.
	In all simulations we observe a slight asymmetry between the $k>0$ and $k<0$ branches, together with a small apparent shift of the cutoff $\kmin$, both of which result from the Lorentz boost in the frame in which the CR shell is isotropic.
	No significant wave power develops at scales that are not resonantly unstable, where $\abs{k} < \kmin$, which corresponds to scales on which power can only be supplied by an inverse cascade.
	
	At both times shown in Fig.~\ref{fig:spectra_baseline}, the LD and NoLD runs agree to within the noise: toggling the Landau closure, and with it NLLD, has no discernible effect on either the growth or the saturation of the self-generated spectrum.
	In particular, the spectral truncation at $k_\mathrm{bal}$ predicted by Eq.~\eqref{eq:kbal} never develops.
	Instead, the spectrum follows the $1/\abs{k}$ shape of the growth rate \eqref{eq:growth_shell} during the growth phase and roughly retains this shape after saturation, consistent with saturation by quasilinear scattering Eq.~\eqref{eq:qlt_spectrum}. A flattening of the spectrum is observed at scales close to $\kmin$, that is linked to the decreasing growth rate starting at around the time $t=5\Gamma_\mathrm{max}^{-1}$ in Fig.~\ref{fig:zpanels}. This is a result of CR bunching that saturates the fastest growing wave modes that are resonant with CRs moving almost parallel ($\mu\approx 1$), see \citet{lemmerz2025} and Appendix~\ref{app:bunching}. Still, growth at smaller modes is not regulated by bunching but quenches once isotropization is reached. This leads to a $1/k$ spectrum consistent with quasilinear isotropization (Eq.~\ref{eq:qlt_spectrum}) in the wave frame $+\valf$, as the spectrum is dominated by $z^-$ modes.

	For a relaxing shell distribution confined by its own waves, NLLD thus plays no role in the saturation.

	\subsection{Low-ppc Simulations: Presence of Spurious Modes}
	\label{sec:res_lowppc}

\begin{figure}[h!]
		\includegraphics[width=\columnwidth]{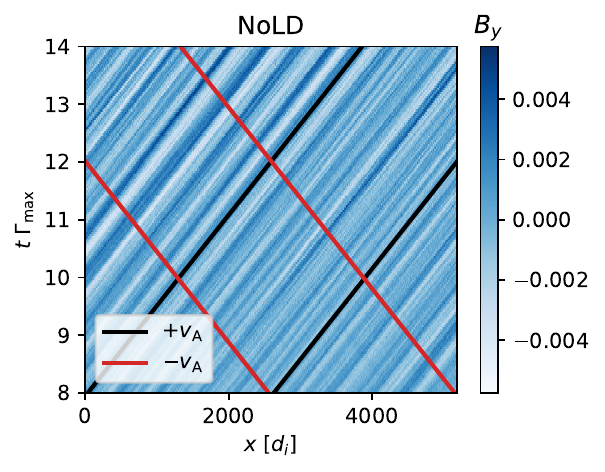}
        
		\includegraphics[width=\columnwidth]{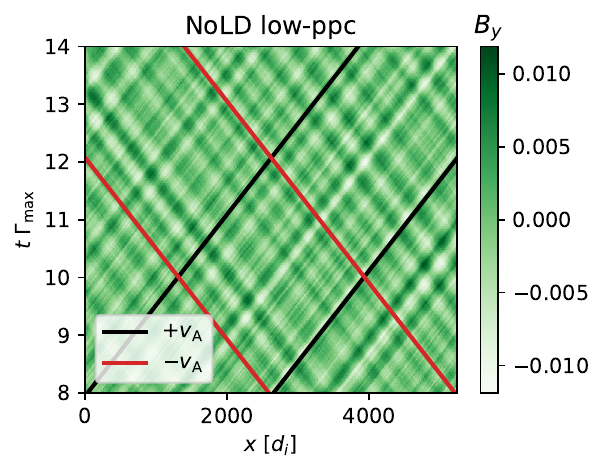}
		\caption{Zoomed-in $x$--$t$ diagram of $B_y(x,t)$ in the \emph{NoLD} (top panel) and \emph{NoLD low-ppc} (bottom panel) simulations. The thick stripes indicate features moving at $+\valf$ (black) and $-\valf$ (red). Both simulations exhibit waves traveling at $+\valf$. Only the low-ppc simulation has additional waves moving at $-\valf$, which should be absent according to linear theory.
		Figure~\ref{fig:zpanels} and Fig.~\ref{fig:mu_res_correlation} show the quantitative decomposition into Elsässer variables over time and $k$, respectively.}
		\label{fig:wk_spectra}
	\end{figure}

\begin{figure*}[t]
		\includegraphics[width=\textwidth]{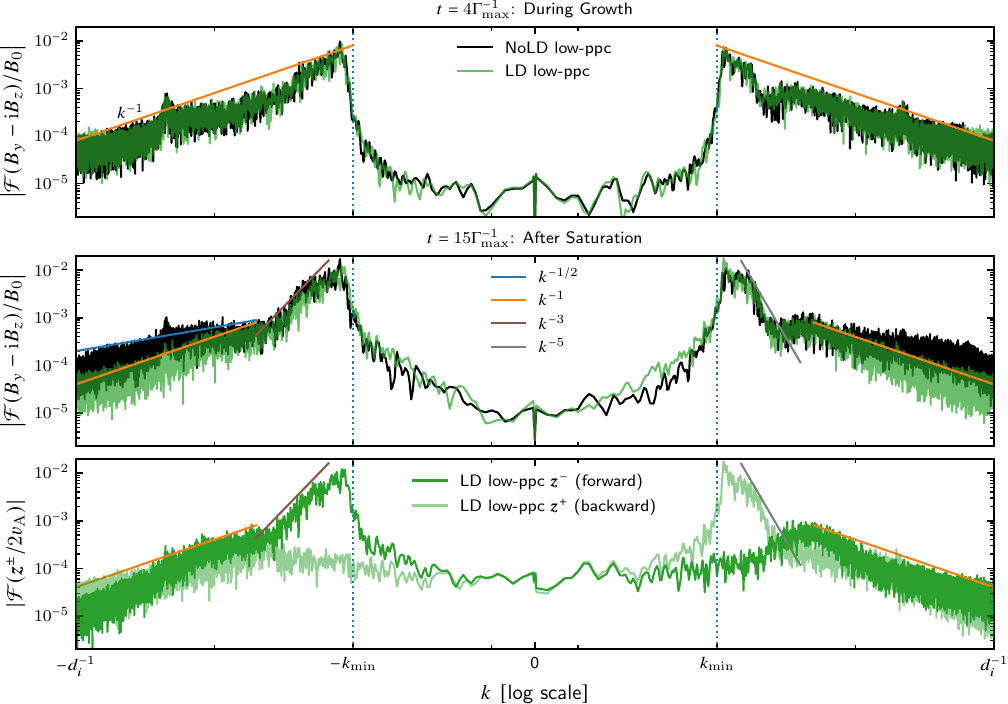}
		\caption{Transverse magnetic spectra of the low-ppc runs with external large-scale modes during growth (top) and after saturation (middle). Reference slopes $k^{-1/2}$, $k^{-1}$, $k^{-3}$ and $k^{-5}$ are indicated.
         A decomposition into Elsässer variables $z^\pm$ of the LD low-ppc simulation is given in the lower panel, whereas the NoLD low-ppc simulation is given in Fig.~\ref{fig:mu_res_correlation} (left panels). Both show significant wave intensity in backward modes close to $\gtrsim+\kmin$, which is not generated by the streaming instability (Eq.~\ref{eq:growth_shell}) and absent from our simulations with more particles-per-cell.
		}
		\label{fig:spectra_lowppc}
	\end{figure*}

	The low-particle-per-cell (low-ppc) runs use twenty times fewer CR macroparticles per $d_i$ than the baseline (Table~\ref{tab:simulations}), which degrades the velocity-space sampling of $f_\crrm(p,\mu)$.
	The consequences on the instability evolution are drastic.
	Figure~\ref{fig:wk_spectra} compares the $x$--$t$ diagrams of the transverse magnetic field. At high particle resolution, all features move forward, as expected for forward moving Alfv\'en waves. In the low-ppc run, however, some features propagate backward, quantified by the high intensity of backward-propagating $z^+$ modes depicted in Fig.~\ref{fig:zpanels}.
	These backward modes are not expected for the streaming instability, but are a result of undersampling the distribution function.
	Yet, these spurious modes may go unnoticed, because the magnetic wave spectra during the growth phase as shown in Fig.~\ref{fig:spectra_lowppc} follow the expected $k^{-1}$ scaling and the wave growth is at a similar level as the expected wave growth.
	Nevertheless, the difference between baseline and low-ppc simulations becomes clear when a decomposition into Els\"asser variables is performed (bottom panel), where the backward moving waves are separated from the forward moving waves.
    Appendix~\ref{app:numerical_instability} performs such an analysis, and investigates the reason for these numerical modes. The poorly sampled CR distribution develops steep, partly \emph{negative} pitch-angle gradients near $\abs{\mu} \to 1$, and these spurious gradients resonantly excite the backward moving modes.

	The macroscopic evolution (Fig.~\ref{fig:zpanels}, low-ppc curves) then deviates from the baseline in four ways.
	Forward waves depart from the linear growth rate early on and grow slightly \emph{faster} than $\Gamma_\mathrm{max}$.
	Resonant backward modes are strongly amplified and eventually reach the forward-mode intensity. As a result of frequent pitch-angle scattering in both wave frames ($+\valf$ and $-\valf$), the drift velocity settles between both frames and approaches $0$, see Section~\ref{sec:drift} for a more in-depth analysis of this.
	Furthermore, the total wave energy does not saturate but keeps growing until the end of the run, because the CRs experience energy diffusion due to frequently scattering with the forward-moving and backward-moving waves in their respective frames. Since the shell distribution is initialized without particles at lower energies, the CR energy gradient $\partial_p f>0$ is positive for some momenta $p$, and thus the energy diffusion refills these low-momentum regions by transferring additional CR energy to the waves.
	
	A remarkable feature of the saturated spectra, as shown in Fig.~\ref{fig:spectra_lowppc}, is a steep power-law spectrum, $B(k) \propto k^{-3}$ for negative $k$ (forward moving waves), or even steeper with positive $k$ ($\abs{k}\gtrsim \kmin$, backward moving waves). Such a steep spectral power law is absent in the baseline and could be misinterpreted as $k_\mathrm{bal}$ given by Eq.~\eqref{eq:kbal}, however, this spectral shape differs for two reasons. First, Eq.~\eqref{eq:kbal} assumes a linearly polarized wave spectrum without backward moving waves, and second, the steep power law exists in both the NoLD and LD simulations, proving that it is independent of NLLD.
	Nevertheless, this steep wave spectrum may still be a result of wave-wave interactions: We point out that the backward moving waves that are concentrated at $\gtrsim +\kmin$ fall off significantly more steeply than $k^{-3}$, and are more in line with a $k^{-5}$ spectrum or even steeper. The $k^{-3}$ power-law spectrum is only observed for the forward moving waves at  $\lesssim -\kmin$.

    Towards larger $k$, the modes are predominantly forward moving ($z^-$) again.
	Without LD, these Alfv\'en modes and the undamped compressible modes driven by the beat waves can still participate in wave-wave interactions, e.g., through parametric decay instabilities, transferring additional energy toward smaller scales. We observe that this flattens the slope of the NoLD spectrum to roughly $k^{-1/2}$ at smaller scales.
	With LD, the energy of the acoustic modes is effectively thermalized, preventing it from being reconverted to Alfv\'en modes. As such, more energy is removed from the smallest scales, leading to an observed steeper spectrum of $k^{-1}$.
	
	Finally, the spectra show growth at non-resonant scales of $\abs{k} < \kmin$ even at late stage -- this is an inverse cascade that is entirely absent in the baseline simulations (Fig.~\ref{fig:zpanels}, bottom right). 
	The strength of this inverse cascade is significantly increased by NLLD, as a result of frequent wave-wave interactions between forward and backward waves close to $k\geq\kmin$. The spurious spectrum is not linearly polarized (as in the baseline), instead, both $z^-$ at $(k<0)$ and $z^+$ at $(k>0)$ rotate left-handedly (cf. Fig.~\ref{fig:convention}) . As a result, the NLLD spectral transfer of Section~\ref{sec:theory_LV} no longer cancels, and power is taken from small-scale modes and either shifted to larger-scale modes or thermalized.
	Similar to the smallest scales, some wave-wave interactions are still possible without NLLD -- however, the inverse cascade happens at a significantly slower rate in the NoLD low-ppc simulation.

	Even though the backward moving modes are generated by spurious numerical effects, their effect on the wave spectrum and on the CRs is physical; in the following section, we inject such waves deliberately and in a controlled way on a scale larger than the maximum resonant wavelength (at $k<\kmin$).
	
	\subsection{+ext Simulations: Damping in Presence of Large-scale External Modes}
	\label{sec:res_ext}
	\begin{figure*}[t]
		\centering\includegraphics[width=\textwidth]{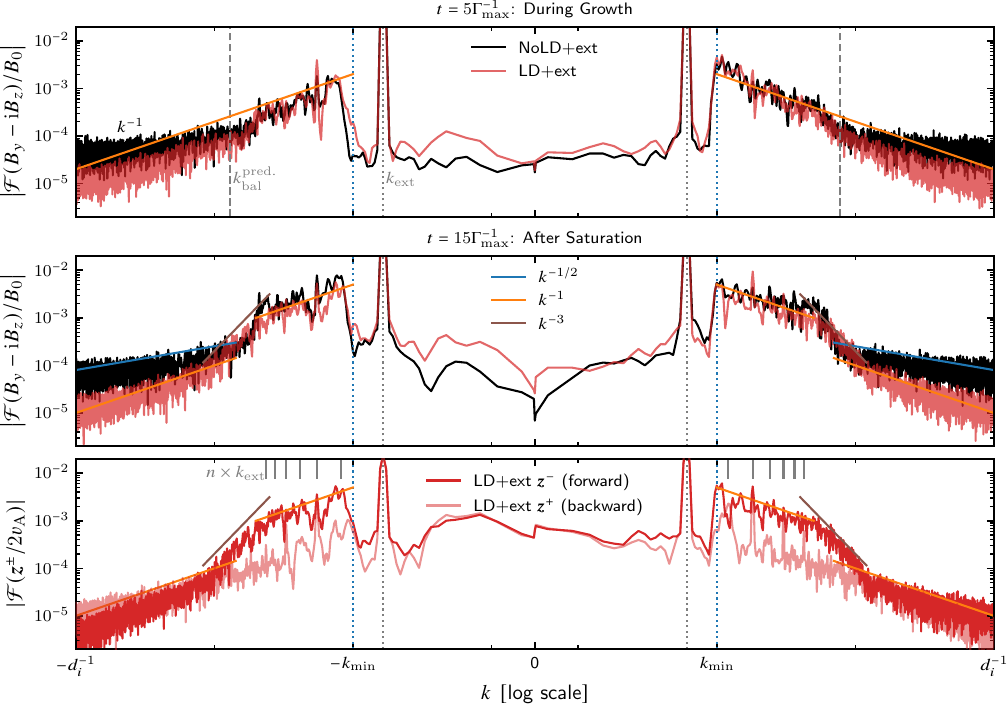}
		\caption{Transverse magnetic spectra of the +ext runs with external large-scale modes during growth (top) and after saturation (middle). The seeded external modes appear as discrete spikes at the grey dotted lines, which are not resonant with the CRs because their $\abs{k} < \kmin$. We also indicate the predicted $k_\mathrm{bal}^\mathrm{LD+ext}\approx7.7\kmin$ as a grey dashed line (Eqs.~\ref{eq:kbal} and~\ref{eq:kbal_ldext}), which marks the scale where streaming and NLLD are expected to balance.  Reference slopes $k^{-1/2}$, $k^{-1}$, and $k^{-3}$ are indicated. The lower panel decomposes the LD+ext spectrum into $z^\pm$, showing that the excited spectrum is predominantly forward moving $(z^-)$, whereas the external modes are balanced. Spikes in the amplitude of $z^+$ are observed at the spatial harmonics, $n\times k_\mathrm{ext}$ for integer values of $n$ between two and seven, indicated with solid grey lines.}
		\label{fig:spectra_ext}
	\end{figure*}

	In the +ext runs, we seed perfectly balanced Alfv\'en modes, namely the four modes $z^+(k_\mathrm{ext})$, $z^+(-k_\mathrm{ext})$, $z^-(k_\mathrm{ext})$ \& $z^-(-k_\mathrm{ext}$), at a scale larger than the maximum CR gyroradius ($k_\mathrm{ext} < \kmin$), with a large amplitude that far exceeds the saturated level of the self-generated waves (dashed line in Fig.~\ref{fig:zpanels}, bottom right).
	Being outside of the gyroresonant range, these modes do not pitch-angle scatter the CRs directly; any influence on the streaming instability must be mediated by wave--wave interactions or mirror-type forces. This is the main difference from the low-ppc type simulations, where backward propagating Alfv\'en modes influence the resonant scales directly.

    Comparing the evolution of the drift velocity and the Els\"asser diagnostics in Fig.~\ref{fig:zpanels} between the baseline and the NoLD+ext run, inclusions of the external modes matter little to the overall evolution. The NoLD+ext run develops the instability somewhat faster, which appears to happen due to an untrapping of the CR bunched particles that are resonant with $\kmin$ (Appendix~\ref{app:bunching}). The NoLD+ext run also overshoots slightly ($\vdr$ drops below $\valf$ before settling at the baseline level just below $\valf$) and shows oscillations in the drift velocity that are in lockstep with the oscillations in the external $B_\perp$ field. These oscillations in the field strength are merely a result of the constructive and destructive interference of forward and backward moving modes, which periodically cancel each other's $B_\perp$ fields -- temporarily converting the magnetic energy to kinetic energy in the background ions $\varv_\perp$ -- before constructively interfering again in $B_\perp$ while canceling out $\varv_\perp$.
	The NoLD+ext simulation shows that the mere presence of large-scale wave power, even if it has a large amplitude, does not significantly affect the macroscopic evolution of the streaming instability if NLLD is turned off. Even though the macroscopic quantities remain similar, the magnetic spectrum still deviates significantly from the baseline, which we will discuss at the end of this section.

	With Landau damping, the evolution changes qualitatively -- and NoLD+ext versus LD+ext is the only pair of runs in which the Landau closure makes a significant difference for the CR drift velocity.
	The forward-wave intensity in LD+ext no longer grows monotonically; instead, NLLD saturates it well below the baseline level.
	Nevertheless, there is enough intensity remaining to scatter the CRs frequently.
	By the end of the simulation, as the intensity of resonant forward and backward moving Alfv\'en waves balance out, the CR drift velocity is observed to drop towards $0$, as expected theoretically (see Section~\ref{sec:drift}). As a result, the LD+ext simulation differs significantly in the final CR distribution from its NoLD+ext counterpart and the baseline simulations.
    However, the resonant $z^+$ waves, whose spectra are illustrated in the lower panel of Fig.~\ref{fig:spectra_ext}, are \emph{not} generated by the CRs. The wave-wave interactions of the external turbulence excites harmonics at scales of $n\times k_\mathrm{ext}$, which are in the resonant regime but grow significantly slower than the streaming instability. While these spikes are also present in the NoLD+ext simulation at a somewhat reduced amplitude, these do not influence the effective CR drift velocity as much due to the large amplitude of forward propagating, $z^+$, modes. The LD+ext simulation damps these modes to an amplitude that is somewhat comparable to the $z^-$ waves by the end of the simulation.

    \begin{figure*}[t]
        \centering
        \includegraphics[width=\textwidth]{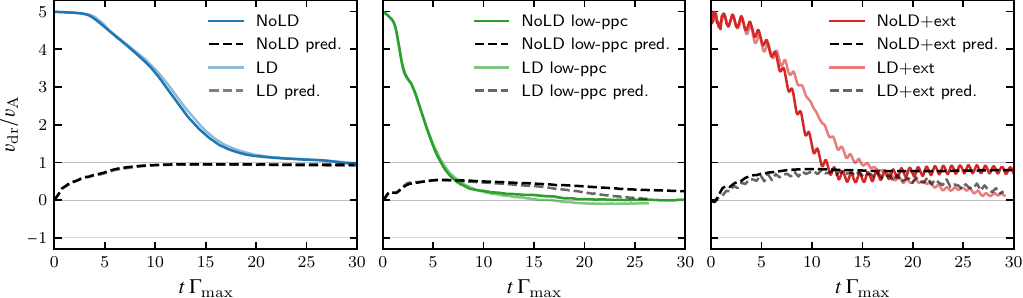}
        \caption{Evolution of the drift velocity measured from the simulations (solid lines) compared with the quasilinear prediction for the steady-state of Eq.~\eqref{eq:effective_drift_velocity} (dashed lines).}
        \label{fig:drift_skilling}
    \end{figure*}
    
	In terms of the magnetic spectra (see Fig.~\ref{fig:spectra_ext}), both +ext simulations deviate from the baseline simulations but are similar to their low-ppc counterparts, and for the same reasons as discussed before. Naively, their magnetic power spectra are similar; yet, the Els\"asser decomposition shows that low-ppc simulations generate significant energy in resonant backward-moving $z^+$ modes, which is absent from the +ext simulations.
	Still, both show a steep power-law spectrum toward large resonant scales and consequent $k^{-1/2}$ (LD+ext) and $k^{-1}$ (NoLD+ext) spectra towards lower $k$. 
    The steep spectra correspond to a $k^{-3}$ power law at both $+k$ and $-k$ modes, which matches low-ppc simulations' spectral shape in the forward moving waves (around $-\kmin$), while their backward moving waves (at $+\kmin$) show a steeper spectrum of $k^{-5}$. This indicates, that the exact power-law index depends on the propagation direction of the Alfv\'en waves.
	The inverse cascade is also present, but significantly reduced in comparison to the low-ppc simulations because the resonant spectra contain less intensity in backward moving wave modes. 
	Unlike all other runs, the +ext simulations show a visible, even if small, reduction in magnetic intensity due to NLLD at the fastest growing resonant wave modes at around $\abs{k}\geq \kmin$.
    It is also the only simulation that shows visible signs of NLLD during the growth phase, because of the presence of pre-seeded waves. The impact during this phase is observed at roughly the same order of magnitude as the estimate of $k_\mathrm{bal}\approx7.7 \kmin$ given in Section~\ref{sec:sim_parameter_ordering}. Yet, the steep power-law that appears at a similar scale after saturation is formed by a physically distinct process, as it also happens in the NoLD simulation.
    This confirms that NLLD \emph{can} make an impact on resonant streaming scales, \emph{if} strong Alfv\'en wave intensity is present at larger scales.
	Yet, if these external waves are present, the excited spectrum is impacted regardless of whether Landau closures are turned on or not, meaning that other processes shape the excited spectrum as well.

	\subsection{Effective Drift Velocity}
        	\label{sec:drift}
    In this section, we link our kinetic results to a formula that may be useful in incorporating subgrid CR physics into galactic simulations. 
	In the quasilinear one-moment approximation, the CR drift speed in the presence of forward- and backward-moving waves is given by the scattering-rate-weighted average of the wave velocities $+ \valf$ and $- \valf$ \citep{skilling1975CosmicRayStreamingI},
	\begin{equation}
		\vdr = \left<\frac{3}{2}\left(1-\mu^2\right)\frac{\nu^F-\nu^B}{\nu^F+\nu^B} \right> \valf,
        \label{eq:effective_drift_velocity}
	\end{equation}
	where $\nu^{F/B}(k_\mathrm{res}(\mu))$ is the quasilinear scattering rate given by Eq.~\eqref{eq:nu_qlt} if only the intensity of forward (F or $z^-$) or backward (B or $z^+$) propagating Alfv\'en waves at $k_\mathrm{res}$ is taken into account and the brackets indicate an average over the CR distribution.
    If only forward moving waves are excited, the CR drift velocity should approach $\valf$ after sufficient time has passed, which is what we see in our baseline simulations. In the presence of only backward moving waves, the drift velocity approaches $-\valf$, and in the presence of both forward and backward moving waves, we expect a drift velocity in between.  In the special case of perfectly balanced scattering rates, $\vdr\to0$ matches the background.

    We apply Eq.~\eqref{eq:effective_drift_velocity} to our simulations. Due to simulation noise in the power spectrum of $B_\perp$, we preprocess the data. The procedure is as follows: We decompose the waves into the Els\"asser variables $z^\pm(k)$, and smooth their spectra using a Gaussian filter with a width of 6 bins in $k$ space, where a bin has length $\Delta k d_i=3\times10^{-4}$ in the baseline and +ext simulations, and a third of that value in the low-ppc simulations. 
    The smoothed spectra is then used to compute the scattering rates $\nu^F (k_\mathrm{res}(\mu))$ and $\nu^B(k_\mathrm{res}(\mu))$. To evaluate the averaging brackets in Eq.~\eqref{eq:effective_drift_velocity} the integral over $\mu$ must be evaluated, $\langle\bullet\rangle \to \int \bullet f_\crrm \mathrm{d}\mu$. Numerically, we use the trapezoidal rule to perform the integration over partitions of length $\Delta\mu=10^{-3}$. To simplify, we roughly approximate the distribution function by $f_\crrm(\mu)\approx \mathrm{const.}$, and calculate the average $u$ from the simulation data at each time step which allows us to find the resonant $k_\mathrm{res} \approx \Omega_i/u \mu$ that is needed to average over $\nu(k_\mathrm{res}(\mu))$.
    
    We find that the effective drift speed in the saturated regime is captured well by Eq.~\eqref{eq:effective_drift_velocity} in the baseline and +ext simulations, as shown in Fig.~\ref{fig:drift_skilling}. Nevertheless, it fails to capture our low-ppc simulations properly. We give two possible explanations. First, the CR population is undersampled and simply behaves differently than it should, see Appendix~\ref{app:numerical_instability}. 
    Second, the low-ppc simulations do not saturate until the end of the simulation and CRs continue to lose a significant amount of energy until the end of the simulation, thereby preventing them from achieving a steady state. It is noteworthy that the power of forward and backward moving waves overall is balanced. As the CRs lose energy, their gyroradii decrease, so Eq.~\eqref{eq:effective_drift_velocity} samples smaller-scale waves that carry comparatively little power. Equation~\eqref{eq:effective_drift_velocity} does not explicitly take into account previous scatterings, but is calculated from the current time slice instead. Only the low-ppc simulations show a large decrease in CR energy, with the velocity at the end of the simulation being roughly half of its initial value, $u\approx 0.5 u_0$, while the other simulations stay at around $97\%$ of the initial value. 

    To conclude, while Eq.~\eqref{eq:effective_drift_velocity} is useful in estimating the CR drift velocity, it is sensitive to the resonant wave spectrum that is unresolved in large-scale galactic simulations. Because of the factor $(1-\mu^2)$, it does not take into account wave energy resonant with CR pitch angles of $\abs{\mu}\approx 1$, which is where most of the generated wave energy is located in our simulations. Instead, wave energy at smaller scales dictates $\vdr$, where the wave spectrum appears to be heavily irregular, specifically in the presence of backward moving waves as shown in Fig.~\ref{fig:spectra_ext}.
    In order to accurately predict the CR drift velocity on a macroscopic scale, it is therefore necessary to also develop physical prescriptions for the resonant wave spectrum that directly influences the effective scattering rate; the effective scattering rate may be simultaneously affected by different processes such as NLLD, pitch angle scattering or gyrophase bunching, complicating the overall picture. 
    
	\section{Discussion}
	\label{sec:discussion}

	\subsection{Implications For More Realistic Cases}
	\label{sec:disc_implications}
	\subsubsection{Difference between Shell and Power-law Distributions}
	\label{sec:disc_powerlaw}
	Our monoenergetic shell distribution provides a clean testbed for investigating NLLD with a specific cutoff at the largest resonant scale $\kmin$, but it is special in one respect: the fastest-growing mode sits directly at the largest unstable scale $\kmin$.
	According to Eq.~\eqref{eq:NLLD_iso}, the fastest-growing mode cannot be damped efficiently in our shell setup, because no larger-scale modes are excited.
	The maximum growth rate between shell and more realistic power-law distributions is similar and even happens at the same scale $\kmin$. Yet, power-law distributions excite waves at larger scales as well. Because NLLD is less efficient at larger scales, the fastest growing wave mode excited by the power-law distributions is not necessarily equivalent to the largest saturated wave mode \citep{hahm2015IonHeatingNonlinear}.
	This allows higher-energy CRs to generate larger scale wave power themselves, which could damp the smaller scales resonant with lower energy CRs (that possess a higher energy density).
	In this context, it is also worth emphasizing that every simulation in which we observed effective NLLD contained counter-propagating waves that can provide more efficient NLLD; a power-law CR population by itself still only drives co-propagating, linearly polarized modes, and the self-consistent NLLD needs to be studied in more detail.
	The cancellation argument of Section~\ref{sec:theory_LV}, and with it the absence of an inverse cascade, applies to the power-law case as well.
    
    Even though investigating NLLD for a drifting power-law CR distribution at first seems like the obvious next step, reality may prove to be even more complex. CR scattering rates are expected to be energy dependent, and thus the drift velocity may be energy dependent as well. Such considerations have been used to fit the observed galactic CR energy spectra to good accuracy \citep{chernyshov2022}.
    Consequently, instead of a power-law distribution with a single drift velocity, a more realistic distribution should allow for an energy dependent $\vdr(p)$. These effects could be investigated numerically using a double-shell distribution with different drift velocities, which would isolate a different aspect of the problem than a power-law with a single-drift velocity.

	\subsubsection{Strength of External Turbulent Energy}
	\label{sec:disc_ext_strength}
	
	Equation~\eqref{eq:NLLD_iso} places no restriction on the scale of external turbulence; the damping rate depends only on the integrated intensity $I_\mathrm{ext}$ over all scales below $k$.
	Taken at face value, this predicts that turbulence at arbitrarily large, e.g.\ galactic, scales damps the streaming modes as efficiently as modes just above the gyroscale -- and, given the large turbulent energy reservoir of the ISM, would quench the streaming instability essentially everywhere.
	Whether the NLLD interaction range in $k$-space really extends this far is, in our view, an open question raised by this work, and we leave it to future study.
	Our +ext runs place the external modes only moderately above the resonant scale, so they demonstrate that large-scale modes \emph{can} curb streaming growth via NLLD, but they do not test the limit of large scale separation.

	We stress that our setup does \emph{not} contain a turbulent cascade.
	Our external modes are parallel-propagating, balanced, and non-cascading in 1D, which isolates the NLLD channel from the anisotropic turbulent damping of \citet{farmer2004WaveDampingMagnetohydrodynamic} \citep[see also][]{yan2002ScatteringCosmicRays,lazarian2021DiffusionCosmicRays,cerri2024}, in which the background cascade shears resonant waves apart, transferring energy toward modes propagating perpendicular to $\bm{B}_0$.
	In the ISM both channels compete, and their relative importance will depend on the amplitude, balance, and anisotropy of the local turbulence -- quantities that are poorly constrained on sub-parsec scales. 
	
	Our results also show that a more realistic estimate of how external turbulence affects NLLD matters, because it directly changes the CR feedback.
	Our LD+ext run isotropizes CRs in the background frame ($\vdr\to0$), allowing them to impart more momentum into the ISM than the NoLD+ext run that isotropizes them in the wave frame ($\vdr\to \valf$). This additional momentum could influence the strength of CR feedback processes such as CR driven winds.
    
    Ultimately, the interaction of CR-driven turbulence with external turbulence remains an open question that is significantly influenced by the specifics of the turbulent cascade in the ISM. 
    Observational evidence points to the ``big power-law in the sky'' \citep{armstrong1995,chepurnov2010} that follows the Kolmogorov turbulence model extending over many decades in $k$-space. Yet this power law may be too simplistic to apply here, because it measure the composite turbulence of many different wave modes. We have, in this paper, only looked at NLLD of incompressible wave modes. 
    Recent simulations find that these modes follow the shallower and more anisotropic Iroshnikov-Kraichnan turbulence \citep{beattie2025}. Nevertheless, even such a unified power-law could be called into question due to local effects introduced by the intermittency of magnetic fields and the multiphase ISM. 
	For now, we conclude that the interaction of external turbulence with self-generated CR streaming modes offers an avenue for further exploration, one that is rich in nonlinear wave--particle and wave--wave interactions.

	\subsubsection{Scaling our Results with Realistic Parameters}
	\label{sec:realistic_scaling}
	Our simulations are designed to keep multiple scale separations related to the CR problem intact: We simulate CR number density ratio  $n_\crrm/n_\mathrm{bg}=5\times10^{-4}$ that is small enough that the growth rate stays below the Alfv\'en wave frequency $\Gamma_\mathrm{max}<\omega(\kmin)$. 
	We also use CRs energetic enough that their speed is well separated from the drift and Alfv\'en speeds, $c\sim \varv\gg\vdr\gtrsim\valf=10^{-2}c$.
	As a result, we do not expect qualitatively new effects by employing ISM parameters, for which streaming growth rates are much smaller due to the minuscule CR density ratio $n_\crrm/n_\mathrm{bg}\sim10^{-9}$ and slower Alfv\'en speed $\valf\sim10^{-4}c$. We do expect, that the saturation amplitude of $B_\perp/B_0$ will be significantly smaller if employing these realistic parameters.

	This raises the question whether NLLD could be quantitatively stronger in the ISM than we found here. However, the NLLD damping rate (Eq.~\ref{eq:NLLD_iso}) and the quasilinear scattering rate (Eq.~\ref{eq:nu_qlt}) both scale essentially with $B^2_\perp/B^2_0$ (although they depend on wave energy at different $k$-scales). Thus, their relative importance for saturation should stay the same when adopting a lower CR density.
	Nevertheless, the gyrophase bunching rate scales significantly differently -- as $B_\perp^{1/2}$ \citep{lemmerz2025} -- so its importance may change in simulations featuring more realistic, slower instability growth compared with our findings in Appendix~\ref{app:bunching}. 
		
	\subsection{Comparison with Other Numerical Studies}
	\label{sec:disc_numerics}
Kinetic simulations of CR streaming that simultaneously resolve the CR gyroscales and the Landau damping of the thermal plasma are computationally challenging, so only a few publications exists so far, which include some with full PIC \citep{holcombMicrophysicsGyroresonantStreaming2019, shalabyNewCosmicRaydriven2020,shalaby2026}, fluid PIC and Landau closure \citep{lemmerz2024CouplingMultifluidDynamics,shalaby2026} and with hybrid-PIC \citep{schroer2025,schroer2026}, of which only the hybrid-PIC studies have reported NLLD as strong enough to curb the instability growth.
	We compare our results with the hybrid-PIC simulations of \citet{schroer2025,schroer2026} (hereafter SCB). The parameter range of \citetalias{schroer2025} is similar to ours, although our CR density is slightly larger.
	Our low-ppc simulations reproduce findings similar to those of \citetalias{schroer2025}, such as the inverse cascade and the quenching of (some) wave growth by NLLD. These low-ppc runs differ substantially from our baseline runs, however, which puts us at odds with their conclusions. If their simulations develop spurious modes like those in our low-ppc runs, the disagreement would be resolved. We therefore briefly examine whether their setup resembles our low-ppc runs more closely than our baseline.

	The number of particles required to prevent the spurious numerical modes found in Section~\ref{sec:res_lowppc} from growing is high: our low-ppc simulation with the lowest CR macroparticle resolution ($204$ CR macroparticles per $d_i$) is comparable in order of magnitude to their simulation (along $\bm{B}_0$ they use $32$ CR macroparticles per $d_i$, but due to a 2D geometry with $10$ cells in the $y$ direction this amounts to roughly $320$ particles in a box of length $d_i\times 5 d_i$). We find in Appendix~\ref{app:numerical_instability} that even our baseline runs with $4080$ particles per $d_i$ experience some minor, unwanted oscillations in the distribution $f_\mathrm{cr}(\mu)$.
	Nevertheless, comparing the number of particles-per-cell is not in itself indicative on whether these spurious modes appear in other simulations -- each code's dynamics and the setup itself are too different to draw conclusions based on that alone. We use a two-fluid background with electron and ion LD, Maxwell's equations and CR particles with high-order shape functions, while \citetalias{schroer2025}'s hybrid-PIC setup uses particles for the background ions only (providing no electron LD) and a generalized Ohm's law. Moreover, the setup is different: we use a ``Lagrangian'' CR streaming setup, where we initially inject CRs and allow them to relax their distribution in response to them driving unstable Alfv\'en waves via the streaming instability, which self-confines the CRs through wave-particle scatterings while \citetalias{schroer2025} use a ``Eulerian'' driven CR streaming setup, where the CRs at the boundaries are continuously resampled, thus maintaining the CR flux and hence, the same instability growth rate. Instead of focusing on the differences in the number of particles and setup, we focus on two points that hint toward a possible presence of the spurious instability in \citetalias{schroer2025}, which potentially explains why they see effects of NLLD comparable to our low-ppc simulations but are at odds with our baseline run.
	
	First, \citet{schroer2026} observe an inverse cascade, which should be absent if only streaming modes were present: the spectral energy transfer from a small-scale to a large-scale wave due to NLLD requires both waves to share the same circular polarization, and for the linearly polarized spectra driven by streaming, the opposing polarization of similar intensity \emph{negates} this transfer by damping at a similar strength, preventing any net energy gain at large scales \citep[see][]{miller1991} and Section~\ref{sec:theory_LV}.
	An inverse cascade therefore signals the presence of waves that streaming alone cannot produce -- as we see in our low-ppc runs (Section~\ref{sec:res_lowppc}).
	Second, \citet{schroer2026} observe a negative gradient in $\mathrm{d} f /\mathrm{d} p_x$ (see their Fig. 3) with steep gradients towards the smallest and largest momenta $p_x$ at the very end of their simulation. While $\mu=p_x/p$ is not shown explicitly, the total momentum $p$ is roughly constant given the monoenergetic initialization, and as a result similar gradients should be observed in $\partial_\mu f$ as well; we identify negative and steep gradients as the cause of the numerical instability in our low-ppc simulations (Appendix~\ref{app:numerical_instability}).

	Even though the rate of NLLD and the importance of the inverse cascade for the streaming instability may have been overstated previously, \citet{schroer2026} main objective was to answer a question that our fluid-PIC code cannot investigate, but that provides valuable insight on whether our method is applicable at all.
	As the effectiveness of Landau damping depends on the slope of the ion distribution function, $\mathrm{d}f_i/\mathrm{d}\varv_x$ at the beat-wave speed, and this slope flattens over time; the effectiveness of NLLD therefore changes as well.
	As a result, the effectiveness of NLLD differs as well. These so-called saturated NLLD damping rates have been previously calculated \citep{volkCesarsky1982, kulsrud1978} and state that damping is provided only by modes local in $k$-space ($\Gamma_\mathrm{sat} \propto k^2 B(k)$), while the unsaturated damping rates depend on the full wave spectrum as given by \citet{miller1991} and Section~\ref{sec:theory_LV}. \citet{lagageCesarsky1983} noted that particle trapping could be unstable in the presence of high amplitude waves and return to the unsaturated damping rate.
	Our Landau closures always damp at the constant, unsaturated rate, and therefore our runs bracket the kinetic damping rates: NoLD runs realize zero NLLD, LD runs realize its theoretical maximum.
	A collisionless kinetic simulation, as performed with hybrid-PIC, should fall in between, depending on the relative rates at which thermal particles are trapped and detrapped.	 
	Consequently, while we use the unsaturated damping rate out of necessity -- our code does not allow for trapping effects in the background distribution -- \citetalias{schroer2025}'s method can differentiate between saturated and unsaturated NLLD. They find that wave energy at substantially larger scales contributes to NLLD as predicted by the unsaturated damping rate; however, a quantitative investigation into how much the unsaturated damping rate is suppressed remains outstanding.

	Regarding possible remedies for the particle-resolution requirement, it has been reported that the $\delta f$ method keeps the necessary particle number affordable in CR streaming setups \citep{bai2019,denton1995DfAlgorithm}, however this comes at the cost of violating particle number and momentum conservation and thereby introduces potentially different numerical effects. More importantly, the $\delta f$ method cannot replicate distribution functions that have zero phase space density at any point, such as the monoenergetic distributions studied here and in \citetalias{schroer2025}, and provides worse results if the CR distribution changes significantly over time; yet, while not always applicable to the CR distribution itself, the $\delta f$ method may be employed to reduce the number of particles in the background distribution function to capture the effect of Landau damping in (hybrid-)PIC simulations \citep{kunz2014}.
	In conventional full $f$ codes, the number of particles per cell remains the main control parameter preventing these spurious modes. However, convergence studies in particle resolution remain impractical due to computational costs. 
	Instead, we recommend comparing the simulated spectra against the theoretical growing wave modes by decomposing the fields into the Elsässer variables, which separates the forward- and backward-propagating Alfv\'en wave modes. Magnetic field arrays are typical outputs of most PIC-derivative codes, but the Elsässer decomposition also requires the background ions' mean $\varv_\perp$ components, which are not routinely saved in some codes. We thus advocate to ensure that this additional information is saved before performing the simulations. 
	
	\section{Conclusion}
	\label{sec:conclusion}

	We have investigated whether nonlinear Landau damping is able to regulate the CR streaming instability, using fluid-PIC simulations in which the Landau closures -- and with them NLLD -- can be switched on and off.
	This direct side-by-side comparison of otherwise identical runs isolates the effect of NLLD in a way that purely kinetic simulations cannot deliver.
	Our CRs are initialized as a monoenergetic shell distribution in a drifting frame, a setup that is unstable to the CR streaming instability.
	We provide four main conclusions.

	First, NLLD of purely self-generated waves is insufficient to quench wave growth for our setup.
	In our baseline runs, the simulations with and without Landau closure are indistinguishable (Section~\ref{sec:res_self}): the instability saturates by bunching (Appendix~\ref{app:bunching}) at the fastest-growing scales and finally by isotropization in the wave frame as the drift velocity settles at $\vdr \simeq \valf$, and the predicted NLLD truncation of the spectrum (Eq.~\ref{eq:kbal}) never develops.
	We expected that NLLD cannot quench the growth of the fastest-growing modes, because the streaming instability generates linearly polarized, co-propagating waves, which NLLD can only damp in the presence of larger-scale wave energy. Such waves are, however, not self-consistently generated by a streaming shell distribution (see Sections~\ref{sec:theory_streaming}), even though power-law distributions may provide some power on larger scales.
	Nevertheless, even the smaller scales that have significantly reduced growth and increased damping rates did not show any difference with NLLD turned on, and thus NLLD was too inefficient to quench growth at any scale.
	Our other sets of simulations differ from this baseline: they contain backward moving Alfv\'en waves that are not generated by the streaming instability, but that substantially alter both NLLD and the resonant CR scattering.	

	Second, we find that simulations of CR streaming at low to moderate macroparticle resolution can develop a numerical instability that qualitatively changes the evolution. Spurious pitch-angle gradients drive backward-propagating Alfv\'en modes, changing the overall dynamics of damping and CR scattering, leading to an isotropization in the background frame (Section~\ref{sec:res_lowppc}).
	Because convergence tests are often impractical at these numerical costs, we recommend monitoring the CR distribution for unphysical gradients and oscillations in $\mathrm{d}f/\mathrm{d}\mu$ -- which may develop at \emph{any} time during a simulation -- and checking the Els\"asser-decomposed spectra against the linear dispersion relation (Section~\ref{sec:disc_numerics} and Appendix~\ref{app:numerical_instability}).
	We emphasize two reasons why this numerical instability may go unnoticed. First, its growth rate is comparable with, though slightly larger than the expected streaming growth rate. Second, because this instability is driven by insufficient velocity-space resolution, it affects the largest resonant scales first. This is unlike spatial-resolution artifacts that typically corrupt the smallest scales first and which are more commonly accounted for. The numerical modes driven by this instability result in a very different late-time evolution.

	Third, we show that high intensity, mixed-propagating Alfv\'en waves -- whether seeded deliberately or introduced inadvertently by insufficient macroparticle resolution -- can effectively quench wave growth at the slower-growing small scales, even though the largest resonant scales remain mostly unaffected (Section~\ref{sec:res_ext}).
	We note that the interaction of some forward- and backward-propagating Alfv\'en modes allows for more efficient NLLD than a purely forward-propagating wave spectrum and that these external waves were seeded at amplitudes larger than the streaming instability provided itself.

	Fourth, in the presence of high intensity external modes, NLLD can change the resonant spectrum by balancing the forward- ($z^-$) and backward- ($z^+$) propagating Alfv\'en waves, which alters CR scattering and leads to isotropization in the background frame -- \emph{not} the Alfv\'en frame -- with $\vdr \ll \valf$, where the drift velocity can be estimated from the balance of the wave spectra (Section~\ref{sec:drift}).
	
	Taken together, our results suggest that NLLD is likely not an intrinsic saturation mechanism of the streaming instability, and that its effectiveness depends crucially on wave power that was not generated by the streaming instability itself.
	For modeling applications, we stress that the turbulent modes of any origin affect the strength of the NLLD regulating CR feedback; for kinetic simulations we stress that numerical instabilities due to insufficient particle resolution may provide that power as well.

	The questions raised by this study directly concern the interaction	of streaming CRs with Alfv\'en waves provided by the turbulent cascade, or possibly even due to pressure-anisotropy type instabilities at larger scales (that would mirror our low-ppc simulations). Whether they provide a viable mechanism in realistic scenarios remains contingent upon the precise parameters of the ISM and further theoretical investigation into the details of the NLLD mechanism.
	On the latter point specifically, the interaction range of NLLD in $k$-space is unknown. Our external modes sit only moderately above the resonant scale, but whether turbulence at much larger, e.g.\ galactic, scales damps streaming modes with undiminished efficiency -- as the theoretical prescriptions predict -- remains to be tested.
	Likewise, the amplitude and balance of ISM turbulence at the sub-parsec scales relevant for CR gyroresonance are difficult to estimate observationally, yet they directly control whether the ISM behaves like our baseline runs, in which CRs stream at the Alfv\'en speed, or like our LD+ext run, in which streaming is partially quenched by NLLD.
	A natural extension to the study provided here is to model CRs with a more realistic power-law distribution. Even though such simulations are computationally demanding because they require high particle resolution and large box sizes to adequately capture the effects of NLLD, their importance for constraining the role of NLLD on CR streaming instabilities can not be overstated.
	
    \section*{Acknowledgments}
    This work was supported in part by the National Science Foundation under Grant No. NSF PHY2409224.
This research was supported in part by Perimeter Institute for Theoretical
Physics. Research at Perimeter Institute is supported by the Government of
Canada through the Department of Innovation, Science and Economic
Development and by the Province of Ontario through the Ministry of Colleges
and Universities. M.S. receives additional support through the Horizon AstroPhysics Initiative (HAPI), a joint venture of the University of Waterloo and Perimeter Institute for Theoretical Physics. CP and TT acknowledge support from the European Research Council under the ERC-AdG grant PICOGAL-101019746. 
This work was supported by the German National High Performance Computing Center G\"ottingen (NHR) under project bbp00078.
	\newpage
	\appendix
	
	\section{Derivation of the Growth Rate of a Monoenergetic, Isotropic CR Distribution}
	\label{app:shell_growth}
	We presume a CR distribution that is monoenergetic and isotropic in some frame,
	\begin{equation}
		f_\mathrm{cr, 0} = n_\crrm \frac{\delta(u-u_0)}{4\pi u^2}, \label{eq:dist_function}
	\end{equation}
	which we will refer to as a shell distribution throughout the paper, as it forms a hollow shell in $u_\parallel, u_\perp$ plots.
	Here, $u=\gamma \varv$ is the proper velocity and $u_0$ is a parameter setting the proper velocity of all CRs.
	We perform a derivation of the growth rate similar to Equation~(23) and onward of \citet{zweibel2003}, using the shell distribution and finally adapting it to the notation introduced in Section~\ref{sec:theory_notation}. The dispersion relation reads as
	\begin{equation}
		\omega^2 - k^2\valf^2 + (\omega-k \vdr)\Omega_i \zeta_\crrm(\abs{k})= 0,
		\label{eq:disp_cr}
	\end{equation}
	where the drift relative to the lab frame is encoded via the term $(\omega-k \vdr)$ and $\vdr\ll c$ is assumed. Moreover, $\omega \ll \Omega_i$ and $\valf \ll c$ are assumed.
    The CR response $\zeta_\crrm$ is symmetric in $k$ due to the assumed isotropy of the CRs in the frame comoving with the initial CR distribution -- half of the population travels backward interacting with $-k$, while the other half travels forward interacting with $+k$ -- and we are interested in its imaginary part given by \citep[compare Equation A9 of][]{zweibel2003}
	\begin{align}
	\Im \zeta_\crrm(\abs{k}) &=  \frac{n_\crrm}{n_i}  \frac{\pi}{2} \abs{u_1} \int_{\abs{u_1}}^{\infty} 4\pi u \frac{f_{\crrm,0}}{n_\crrm} \mathrm{d} u\\
	\overset{\text{shell}}{\Longrightarrow} \Im \zeta_\crrm(\abs{k})&= \frac{n_\crrm}{n_i}  \frac{\pi}{2} \frac{\abs{u_1}}{u_0} \Theta(u_0-u_1), \text{ where } u_1=\abs{\frac{\Omega_i}{k}}
	\end{align}
	The velocity $u_1$ is the minimum velocity required for gyroresonance with a wavenumber $k$, under the approximation of $\valf \ll \Omega_i/k$ (Eq.~\ref{eq:res_approx}). 
	The Heaviside function $\Theta$ expresses that $u_0$ must be larger than this minimum velocity $u_1(k)$ to interact resonantly with a mode at scale $k$.
	The solution of Eq.~\eqref{eq:disp_cr} is, up to first order in the small parameter $\zeta_\crrm\sim n_\crrm/n_i$,
	\begin{align}
		\omega(k) =& \pm\sign(k) k \valf  +  \frac{\zeta_\crrm(\abs{k} )}{2}\left(-1\pm\sign(k) \frac{\vdr}{\valf} \right) \Omega_i  + \mathcal{O}(\zeta_\crrm^2), 
        \end{align}
        \begin{align}
		\Gamma(k) = &\Im(\omega(k)) = \frac{n_\crrm}{n_i} \frac{\pi}{4} \abs{\frac{\Omega_i}{k}}\frac{1}{u_0} \Theta(u_0-u_1)  \times \left(-1\pm\sign(k) \frac{\vdr}{\valf} \right) \Omega_i+ \mathcal{O}(\zeta_\crrm^2) . \label{eq:Growth_rate}
	\end{align}
	We note that $\pm$ in this equation does not denote the wave propagation direction as elsewhere in this paper, but has no clear physical interpretation. To rewrite this in terms of propagation direction, we find that the only modes that can grow given $\vdr>\valf$ must satisfy $\pm \sign(k)>0$.
	These modes fulfill $\omega=+k \valf$ (to zeroth order in $\zeta_\crrm$) and thus are forward propagating waves or $z^-$ modes; backward propagating $z^+$ modes are damped. We adapt the sign convention correspondingly to arrive at Eq.~\eqref{eq:growth_shell}.
	
	\section{A Numerical Instability Caused by Insufficient Particle Resolution}
	\label{app:numerical_instability}
	
		\begin{figure*}[h!]
	    \includegraphics[width=.49\columnwidth]{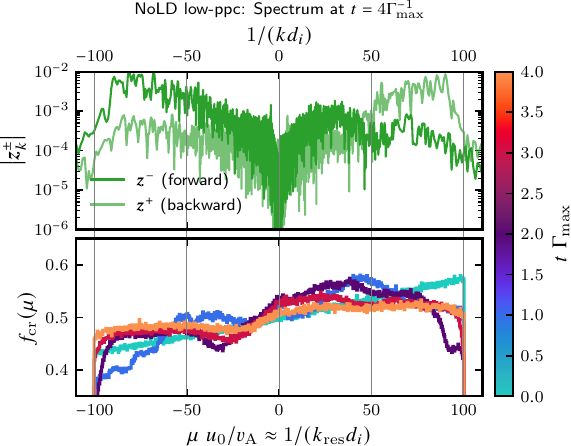}
	    \includegraphics[width=.49\columnwidth]{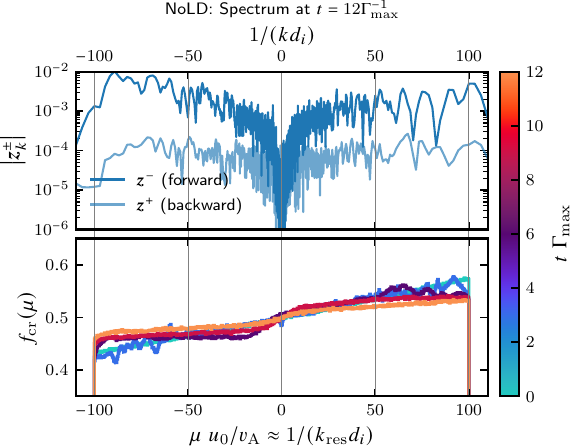}
		\caption{Insufficient CR particle resolution leads to unphysical gradients in $\mathrm{d}f/\mathrm{d}\mu$ that drive numerical instabilities. The top panels compare the spectra of the low-ppc (left) and baseline (right) runs as each simulation approaches $\vdr\approx 3\valf$, i.e., corresponding to the maximum time shown in the bottom panels (orange in the color scale). The bottom panels show the time evolution of $f_\crrm(\mu)$, i.e. the distributions that excited the spectra above. The $x$-axis is scaled by $u_0/\valf=100$ so that it matches the resonant $k$ in the top panel. 
		Left panels:
		In the low-ppc simulation, the CR distribution function is significantly distorted towards $\abs{\mu} \lessapprox 1$, generating spurious modes as $\partial_{\mu} f_\crrm(\mu)$ shows steeper, sometimes even negative, gradients. The negative gradients lead to a generation of backward moving Alfv\'en wave modes ($z^+$).
		Right panels: the streaming instability predominantly generates forward propagating Alfv\'en waves, because the CR distribution is sufficiently well sampled and $\partial_{\mu} f_\crrm(\mu)$ stays (mostly) positive, except for some minor oscillations.}
		\label{fig:mu_res_correlation}
	\end{figure*}

	The PIC method represents the CR distribution function by a finite number of macroparticles, so $f_\crrm(p,\mu)$ is subject to Poisson sampling noise whose relative amplitude scales inversely with the square root of the number of particles per resonant interval. The number of macroparticles is directly linked to the velocity-space resolution and how well $f_\crrm(p,\mu)$ is sampled. In CR streaming simulations, the CRs carry a significant fraction of the overall energy that is present in the simulation box, so undersampling them can lead to a drastically overestimated energy deposition.
	In the low-ppc runs, which use twenty times fewer CR macroparticles than the baseline, the undersampled and noisy $f_\crrm(\mu)$ distribution develops steep gradients in $\partial_\mu f_\crrm$ that are very different from the initialized drifting-shell distribution function.
	
	Although these spurious gradients are a sampling artifact, the waves they excite are a \emph{physically correct} consequence; the electromagnetic field solver and the background plasma respond correctly to the (wrong) CR distribution they are presented with.
	The sign convention of Section~\ref{sec:theory_notation} helps to understand this response.
	A CR with $\mu > 0$ is gyroresonant with modes at $k > 0$ and a CR with $\mu < 0$ with modes at $k < 0$, \emph{independent of the propagation direction} of the wave, see Eq.~\eqref{eq:res_theory} and Fig.~\ref{fig:convention}. So a backward wave $z^+(k)$ with $k>0$ resonates with the same CRs as a forward wave $z^-(k)$ with $k>0$.
	Which of the two propagation directions grows is decided by the sign of the pitch-angle gradient at the resonant $\mu$. The drifting shell has a strictly \emph{positive} $\partial_\mu f_\crrm > 0$ when initialized, which amplifies only forward ($z^-$) modes and damps backward ($z^+$) modes (Appendix~\ref{app:shell_growth}), whereas a locally \emph{negative} gradient inverts the sign of the resonant CR response and drives the backward modes unstable.
	The spurious backward waves also couple back onto the CRs in a physically consistent way. Once they have grown exponentially, they scatter the CRs as genuine waves would, with the macroscopic consequences described in Section~\ref{sec:res_lowppc}.

	Figure~\ref{fig:mu_res_correlation} demonstrates this correlation between velocity-space gradients and wave content directly, comparing the low-ppc and baseline runs at the same stage ($\vdr \approx 3\valf$).
	The $x$-axes of the top and bottom panels are matched through the resonance condition, so wave power at a given $k$ appears directly above the pitch angle $\mu$ of the CRs that drive it, albeit the wave power will lag behind in time as the instability develops -- which is why we plot the time evolution of $f_\crrm(\mu)$ leading up to the spectrum.
	In the low-ppc run, the regions of distorted, partly negative $\partial_\mu f_\crrm$ during its time evolution coincide with substantial backward ($z^+$) wave power, while in the baseline run the smooth, (almost) monotonically increasing $f_\crrm(\mu)$ generates mostly forward ($z^-$) modes. These spurious modes are concentrated towards the fastest growing modes resonant with $\abs{\mu} \lesssim 1$ (Fig.~\ref{fig:mu_res_correlation}, bottom left), and the larger $\partial_\mu f_\crrm$ gradients also lead to faster growth overall, as visible in Fig.~\ref{fig:zpanels}.
	Nevertheless, even the well-resolved baseline shows minor oscillations in $f_\crrm(\mu)$; the sampling artifact is suppressed by particle number and not outright eliminated. Still, the suppression is good enough to prevent the build-up of $z^+$ modes to any appreciable degree.

	This instability is easy to miss for two reasons.
	First, the spurious gradients are transient: once the CRs are well scattered, pitch-angle diffusion smooths $f_\crrm(\mu)$ and the tell-tale distortions largely disappear, so they can only be detected during the early stages of a simulation. Nevertheless, the backward waves they excited persist and influence the spectrum, CR energization, and saturation until the very end of the run (Section~\ref{sec:res_lowppc}).
	Second, the instability affects the largest \emph{resonant} scales, around $\kmin$, where the growth rates peak.
	Numerical artifacts are expected at the smallest resolved scales, as is the case for spatial-resolution artifacts, so an artifact like this that appears directly in the regime of interest, well separated from both the simulation box and step sizes, may be easily mistaken for a physical feature.
	The diagnostics that we recommend in Section~\ref{sec:disc_numerics} include specifically monitoring $\mathrm{d}f/\mathrm{d}\mu$ for unphysical gradients at a high cadence during the exponential growth phase and at lower cadence throughout the run, and checking the Els\"asser-decomposed $z^\pm(k)$ spectra against the linear dispersion relation.

	\section{Saturation due to Gyrophase Bunching}
	\label{app:bunching}
	\begin{figure*}[h]
	\includegraphics[width=.49\textwidth]{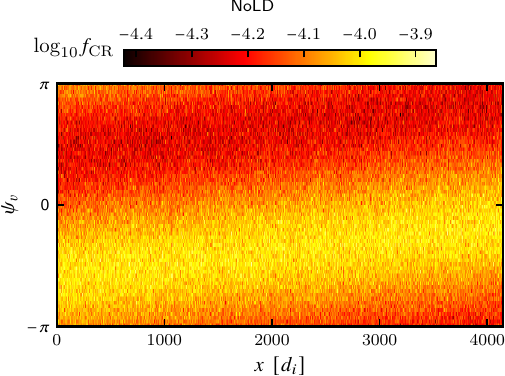}
	\includegraphics[width=.49\textwidth]{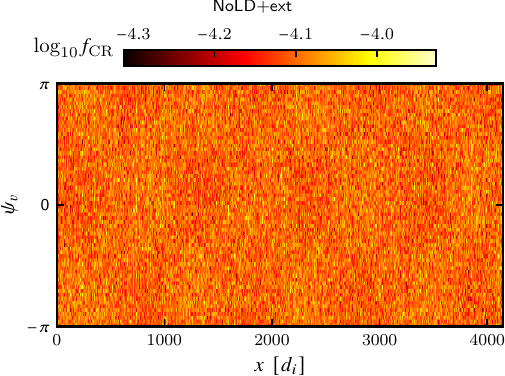}
	\caption{Gyrophase angle histograms ($\psi_\varv=\angle(\varv_y-\ci \varv_z)$) during growth at $t=6~\Gamma_\mathrm{max}^{-1}$, which is when the instability growth of the NoLD (and LD) simulation falls behind that of NoLD+ext (see Fig.~\ref{fig:zpanels}). We plot only the particles with $\mu>0.7$ that are resonantly driving the fastest growing wave modes in a subsection of the box. NoLD shows a gyrophase bunching signature that is absent in the +ext counterpart, where particles are untrapped due to the large external intensity.}
	\label{fig:gyrophase_bunching}
\end{figure*}	
	In the baseline runs, the growth of the resonant forward waves falls behind the linear expectation at $t \approx 5$--$6\,\Gamma_\mathrm{max}^{-1}$ (Fig.~\ref{fig:zpanels}, top right), even though the drift velocity -- and with it the predicted linear growth rate -- is still close to its initial value at that time.
	Because this early departure occurs identically with and without the Landau closure, it cannot be caused by (nonlinear) Landau damping.
	We attribute it to \emph{gyrophase bunching} of the resonant CRs, a saturation mechanism of resonant electromagnetic instabilities \citep{sudan1971TheoryTriggeredVLF} with applications to resonant Alfv\'en-wave beam instabilities  \citep{otani1988ApplicationNonlinearDynamical,zachary1989LongtimeEvolutionLowdensity} and, more specifically, CR instabilities \citep{lemmerz2025, shalaby2026}; here we summarize the mechanism and show how it links to the simulations at hand.

	A CR satisfying the resonance condition \eqref{eq:res_theory} rotates in phase with the perpendicular field of the resonant Alfv\'en wave, so the energy and pitch-angle change of the particle is controlled by the relative angle between its perpendicular velocity and the local wave field. At resonance, this angle remains roughly constant for a long time.
	Linear growth implicitly assumes that the resonant CRs remain uniformly distributed in this angle, sampling all phases of the wave.
	As the wave amplitude grows, however, near-resonant particles become trapped in the wave field and oscillate, stalling the exponential growth.
	
	Figure~\ref{fig:gyrophase_bunching} shows this bunching directly.
	We plot the phase-space density of the CRs with $\mu > 0.7$ -- the population that resonantly drives the fastest-growing modes at the large-scale end of the spectrum, $\abs{k} \gtrsim \kmin$ -- as a function of position and gyrophase $\psi_\varv$, at $t = 6\,\Gamma_\mathrm{max}^{-1}$, the time at which the baseline growth falls behind that of NoLD+ext.
	In the NoLD baseline, the resonant CR density is clearly modulated in gyrophase, coherently across the simulation box.
	In NoLD+ext, it is uniform, as the large-amplitude external modes continuously perturb the resonance and untrap the particles.

	This explains the difference in growth rates between these simulations in Fig.~\ref{fig:zpanels}.
	In the baseline runs nothing disturbs the resonances, so the fastest-growing modes trap their resonant particles and saturate early.
	Gyrophase bunching is thus the first stage of the two-fold saturation described in Section~\ref{sec:res_self} and regulates the growth of the fastest growing and most intense wave modes. Waves at larger $k$, which carry less wave intensity, instead saturate as the CRs isotropize due to frequent QLT pitch-angle scattering.
	The NoLD+ext run keeps its resonant CRs untrapped and sustains growth close to the linear rate for longer (Section~\ref{sec:res_ext}).

\bibliography{lib}

@ARTICLE{pfrommer2013,
       author = {{Pfrommer}, Christoph},
        title = "{Toward a Comprehensive Model for Feedback by Active Galactic Nuclei: New Insights from M87 Observations by LOFAR, Fermi, and H.E.S.S.}",
      journal = {\apj},
         year = 2013,
        month = dec,
       volume = {779},
       number = {1},
          eid = {10},
        pages = {10},
          doi = {10.1088/0004-637X/779/1/10},
archivePrefix = {arXiv},
       eprint = {1303.5443},
 primaryClass = {astro-ph.CO},
       adsurl = {https://ui.adsabs.harvard.edu/abs/2013ApJ...779...10P}
}

@ARTICLE{shalaby2026,
       author = {{Shalaby}, Mohamad and {Lemmerz}, Rouven and {Pfrommer}, Christoph},
        title = "{Saturation Mechanism of Cosmic Ray Streaming Instabilities with a Shell Distribution}",
      journal = {arXiv e-prints},
         year = 2026,
        month = sep,
          eid = {arXiv:2609.30376},
        pages = {arXiv:2609.30376},
          doi = {10.48550/arXiv.2609.30376},
archivePrefix = {arXiv},
       eprint = {2609.30376},
 primaryClass = {astro-ph.GA},
       adsurl = {https://ui.adsabs.harvard.edu/abs/2026arXiv260930376S}
}

@article{bliokh2022,
  title = {Momentum, Angular Momentum, and Spin of Waves in an Isotropic Collisionless Plasma},
  author = {Bliokh, Konstantin Y. and Bliokh, Yury P.},
  year = 2022,
  month = jun,
  journal = {Physical Review E},
  volume = {105},
  number = {6},
  pages = {065208},
  publisher = {American Physical Society},
  doi = {10.1103/PhysRevE.105.065208},
  urldate = {2026-04-23}
}

@article{beattie2025,
  title = {So {{Long Kolmogorov}}: {{The Forward}} and {{Backward Turbulence Cascades}} in a {{Supernovae-driven}}, {{Multiphase Interstellar Medium}}},
  shorttitle = {So {{Long Kolmogorov}}},
  author = {Beattie, James R. and Kolborg, Anne Noer and {Ramirez-Ruiz}, Enrico and Federrath, Christoph},
  year = 2025,
  month = dec,
  journal = {The Astrophysical Journal},
  volume = {994},
  pages = {193},
  publisher = {IOP},
  issn = {0004-637X},
  doi = {10.3847/1538-4357/ae07cd},
  urldate = {2026-09-11}
}

@article{blasi2012,
  title = {Spectral {{Breaks}} as a {{Signature}} of {{Cosmic Ray Induced Turbulence}} in the {{Galaxy}}},
  author = {Blasi, Pasquale and Amato, Elena and Serpico, Pasquale D.},
  year = 2012,
  month = aug,
  journal = {Physical Review Letters},
  volume = {109},
  number = {6},
  pages = {061101},
  publisher = {American Physical Society},
  doi = {10.1103/PhysRevLett.109.061101},
  urldate = {2024-12-07}
}

@book{kulsrud2004PlasmaPhysicsAstrophysics,
  title = {Plasma {{Physics}} for {{Astrophysics}}},
  author = {Kulsrud, Russell},
  year = 2004,
  publisher = {Princeton University Press},
  urldate = {2021-04-01},
  isbn = {978-0-691-12073-7},
  langid = {english}
}

@article{xu2022,
  title = {Cosmic {{Ray Streaming}} in the {{Turbulent Interstellar Medium}}},
  author = {Xu, Siyao and Lazarian, Alex},
  year = 2022,
  month = mar,
  journal = {The Astrophysical Journal},
  volume = {927},
  pages = {94},
  publisher = {IOP},
  issn = {0004-637X},
  doi = {10.3847/1538-4357/ac4dfd},
  urldate = {2026-09-17}
}

@article{thomas2026,
  title = {Effective Cosmic-Ray Diffusion in Multiphase Galactic Environments},
  author = {Thomas, T. and Pfrommer, C. and Pakmor, R. and Lemmerz, R. and Shalaby, M.},
  year = 2026,
  month = aug,
  journal = {Astronomy and Astrophysics},
  volume = {712},
  pages = {A209},
  publisher = {EDP},
  issn = {0004-6361},
  doi = {10.1051/0004-6361/202557765},
  urldate = {2026-09-17}
}

@article{zweibel2017,
  title = {The Basis for Cosmic Ray Feedback: {{Written}} on the Wind},
  shorttitle = {The Basis for Cosmic Ray Feedback},
  author = {Zweibel, Ellen G.},
  year = 2017,
  month = may,
  journal = {Physics of Plasmas},
  volume = {24},
  number = {5},
  pages = {055402},
  publisher = {American Institute of Physics},
  issn = {1070-664X},
  doi = {10.1063/1.4984017},
  urldate = {2022-03-17}
}

@article{ruszkowskipfrommer2023,
  title = {Cosmic Ray Feedback in Galaxies and Galaxy Clusters: {{A}} Pedagogical Introduction and a Topical Review of the Acceleration, Transport, Observables, and Dynamical Impact of Cosmic Rays},
  shorttitle = {Cosmic Ray Feedback in Galaxies and Galaxy Clusters},
  author = {Ruszkowski, Mateusz and Pfrommer, Christoph},
  year = 2023,
  month = dec,
  journal = {The Astronomy and Astrophysics Review},
  volume = {31},
  number = {1},
  pages = {4},
  issn = {0935-4956, 1432-0754},
  doi = {10.1007/s00159-023-00149-2},
  urldate = {2024-10-02},
  langid = {english}
}

@article{xie2014PDRFGeneralDispersion,
  title = {{{PDRF}}: {{A}} General Dispersion Relation Solver for Magnetized Multi-Fluid Plasma},
  shorttitle = {{{PDRF}}},
  author = {Xie, Hua-sheng},
  year = 2014,
  month = feb,
  journal = {Computer Physics Communications},
  volume = {185},
  number = {2},
  pages = {670--675},
  issn = {0010-4655},
  doi = {10.1016/j.cpc.2013.10.012},
  urldate = {2020-07-20},
  langid = {english}
}

@article{hopkins2022,
  title = {Standard Self-Confinement and Extrinsic Turbulence Models for Cosmic Ray Transport Are Fundamentally Incompatible with Observations},
  author = {Hopkins, Philip F. and Squire, Jonathan and Butsky, Iryna S. and Ji, Suoqing},
  year = 2022,
  month = dec,
  journal = {Monthly Notices of the Royal Astronomical Society},
  volume = {517},
  pages = {5413--5448},
  publisher = {OUP},
  issn = {0035-8711},
  doi = {10.1093/mnras/stac2909},
  urldate = {2026-09-17}
}

@article{recchia2022,
  title = {Grammage of Cosmic Rays in the Proximity of Supernova Remnants Embedded in a Partially Ionized Medium},
  author = {Recchia, S. and Galli, D. and Nava, L. and Padovani, M. and Gabici, S. and Marcowith, A. and Ptuskin, V. and Morlino, G.},
  year = 2022,
  month = apr,
  journal = {Astronomy and Astrophysics},
  volume = {660},
  pages = {A57},
  publisher = {EDP},
  issn = {0004-6361},
  doi = {10.1051/0004-6361/202142558},
  urldate = {2026-09-17}
}

@article{chernyshov2022,
	title = {Formation of the {{Cosmic-Ray Halo}}: {{The Role}} of {{Nonlinear Landau Damping}}},
	shorttitle = {Formation of the {{Cosmic-Ray Halo}}},
	author = {Chernyshov, D. O. and Dogiel, V. A. and Ivlev, A. V. and Erlykin, A. D. and Kiselev, A. M.},
	year = 2022,
	month = oct,
	journal = {The Astrophysical Journal},
	volume = {937},
	number = {2},
	pages = {107},
	issn = {0004-637X, 1538-4357},
	doi = {10.3847/1538-4357/ac8f42},
	urldate = {2026-05-18}
}

@article{chernyshov2024,
  title = {Secondary Cosmic-Ray Nuclei in the {{Galactic}} Halo Model with Nonlinear {{Landau}} Damping},
  author = {Chernyshov, D. O. and Ivlev, A. V. and Dogiel, V. A.},
  year = 2024,
  month = jun,
  journal = {Astronomy and Astrophysics},
  volume = {686},
  pages = {A165},
  publisher = {EDP},
  issn = {0004-6361},
  doi = {10.1051/0004-6361/202348766},
  urldate = {2026-09-17}
}

@article{sike2025,
  title = {Cosmic-{{Ray-driven Galactic Winds}} with {{Resolved Interstellar Medium}} and {{Ion-neutral Damping}}},
  author = {Sike, Brandon and Thomas, Timon and Ruszkowski, Mateusz and Pfrommer, Christoph and Weber, Matthias},
  year = 2025,
  month = jul,
  journal = {The Astrophysical Journal},
  volume = {987},
  pages = {204},
  publisher = {IOP},
  issn = {0004-637X},
  doi = {10.3847/1538-4357/adda3d},
  urldate = {2026-09-17}
}

@article{aloisio2013,
  title = {Propagation of Galactic Cosmic Rays in the Presence of Self-Generated Turbulence},
  author = {Aloisio, Roberto and Blasi, Pasquale},
  year = 2013,
  month = jul,
  journal = {Journal of Cosmology and Astroparticle Physics},
  volume = {2013},
  pages = {001},
  publisher = {IOP},
  issn = {1475-7516},
  doi = {10.1088/1475-7516/2013/07/001},
  urldate = {2024-12-12}
}

@article{kempski2022,
  title = {Reconciling Cosmic Ray Transport Theory with Phenomenological Models Motivated by {{Milky-Way}} Data},
  author = {Kempski, Philipp and Quataert, Eliot},
  year = 2022,
  month = jul,
  journal = {Monthly Notices of the Royal Astronomical Society},
  volume = {514},
  number = {1},
  pages = {657--674},
  issn = {0035-8711},
  doi = {10.1093/mnras/stac1240},
  urldate = {2024-11-19}
}

@article{shalaby2017,
  title = {{{SHARP}}: {{A Spatially Higher-order}}, {{Relativistic Particle-in-cell Code}}},
  shorttitle = {{{SHARP}}},
  author = {Shalaby, Mohamad and Broderick, Avery E. and Chang, Philip and Pfrommer, Christoph and Lamberts, Astrid and Puchwein, Ewald},
  year = 2017,
  month = may,
  journal = {The Astrophysical Journal},
  volume = {841},
  number = {1},
  pages = {52},
  publisher = {American Astronomical Society},
  issn = {0004-637X},
  doi = {10.3847/1538-4357/aa6d13},
  urldate = {2020-10-23},
  langid = {english}
}

@article{andrews1978,
  title = {On Wave-Action and Its Relatives},
  author = {Andrews, D. G. and Mcintyre, M. E.},
  year = 1978,
  month = dec,
  journal = {Journal of Fluid Mechanics},
  volume = {89},
  number = {4},
  pages = {647--664},
  issn = {0022-1120, 1469-7645},
  doi = {10.1017/S0022112078002785},
  urldate = {2026-09-10},
  copyright = {https://www.cambridge.org/core/terms},
  langid = {english}
}

@article{kulsrud1969,
  author =        {Kulsrud, Russell and Pearce, William P.},
  journal =       {The Astrophysical Journal},
  month =         may,
  pages =         {445},
  title =         {The {{Effect}} of {{Wave-Particle Interactions}} on
                   the {{Propagation}} of {{Cosmic Rays}}},
  volume =        {156},
  year =          {1969},
  doi =           {10.1086/149981},
  issn =          {0004-637X, 1538-4357},
}

@article{hollweg1971NonlinearLandauDamping,
  author =        {Hollweg, Joseph V.},
  journal =       {Physical Review Letters},
  month =         nov,
  number =        {20},
  pages =         {1349--1352},
  title =         {Nonlinear {{Landau Damping}} of {{Alfv\'en Waves}}},
  volume =        {27},
  year =          {1971},
  doi =           {10.1103/PhysRevLett.27.1349},
  issn =          {0031-9007},
}

@article{loewenstein1991,
  title = {Cosmic-{{Ray Heating}} of {{Cooling Flows}}: {{A Critical Analysis}}},
  shorttitle = {Cosmic-{{Ray Heating}} of {{Cooling Flows}}},
  author = {Loewenstein, Michael and Zweibel, Ellen G. and Begelman, Mitchell C.},
  year = 1991,
  month = aug,
  journal = {The Astrophysical Journal},
  volume = {377},
  pages = {392},
  publisher = {IOP},
  issn = {0004-637X},
  doi = {10.1086/170369},
  urldate = {2026-10-08}
}

@article{hollweg1971DensityFluctuationsDriven,
  title = {Density Fluctuations Driven by {{Alfv\'en}} Waves},
  author = {Hollweg, Joseph V.},
  year = 1971,
  month = aug,
  journal = {Journal of Geophysical Research},
  volume = {76},
  number = {22},
  pages = {5155--5161},
  issn = {01480227},
  doi = {10.1029/JA076i022p05155},
  urldate = {2021-05-07},
  langid = {english}
}

@book{stix1992,
  title = {Waves in Plasmas},
  author = {Stix, Thomas Howard},
  year = 1992,
  publisher = {AIP},
  address = {New York},
  isbn = {978-0-88318-859-0},
  langid = {english}
}

@article{lee1973DampingNonlinearWaveparticle,
  author =        {Lee, Martin A. and V{\"o}lk, Heinrich J.},
  journal =       {Astrophysics and Space Science},
  month =         sep,
  number =        {1},
  pages =         {31--49},
  title =         {Damping and Nonlinear Wave-Particle Interactions of
                   {{Alfv\'en-waves}} in the Solar Wind},
  volume =        {24},
  year =          {1973},
  doi =           {10.1007/BF00648673},
  issn =          {1572-946X},
}

@article{volkCesarsky1982,
  author =        {V{\"o}lk, Heinrich J. and Cesarsky, Catherine J.},
  chapter =       {Zeitschrift f\"ur Naturforschung A},
  journal =       {Zeitschrift f\"ur Naturforschung A},
  month =         aug,
  number =        {8},
  pages =         {809--815},
  publisher =     {Verlag der Zeitschrift f\"ur Naturforschung},
  title =         {Nonlinear {{Landau Damping}} of {{Alfv\'en Waves}} in
                   a {{High}} Beta {{Plasma}}},
  volume =        {37},
  year =          {1982},
  doi =           {10.1515/zna-1982-0814},
  issn =          {1865-7109},
}

@article{miller1991,
  author =        {Miller, James A.},
  journal =       {The Astrophysical Journal},
  month =         jul,
  pages =         {342},
  title =         {Magnetohydrodynamic Turbulence Dissipation and
                   Stochastic Proton Acceleration in Solar Flares},
  volume =        {376},
  year =          {1991},
  doi =           {10.1086/170284},
  issn =          {0004-637X, 1538-4357},
}

@article{ptuskin2008NonlinearDiffusionCosmic,
  author =        {Ptuskin, V. S. and Zirakashvili, V. N. and
                   Plesser, A. A.},
  journal =       {Advances in Space Research},
  month =         aug,
  number =        {3},
  pages =         {486--490},
  title =         {Non-Linear Diffusion of Cosmic Rays},
  volume =        {42},
  year =          {2008},
  doi =           {10.1016/j.asr.2007.12.007},
  issn =          {0273-1177},
}

@article{guo2008,
  title = {Feedback Heating by Cosmic Rays in Clusters of Galaxies},
  author = {Guo, Fulai and Oh, S. Peng},
  year = 2008,
  month = feb,
  journal = {Monthly Notices of the Royal Astronomical Society},
  volume = {384},
  number = {1},
  pages = {251--266},
  issn = {0035-8711},
  doi = {10.1111/j.1365-2966.2007.12692.x},
  urldate = {2026-09-29}
}

@article{wiener2013CosmicRayStreaming,
  author =        {Wiener, Joshua and Oh, S. Peng and Guo, Fulai},
  journal =       {Monthly Notices of the Royal Astronomical Society},
  month =         sep,
  pages =         {2209--2228},
  publisher =     {OUP},
  title =         {Cosmic Ray Streaming in Clusters of Galaxies},
  volume =        {434},
  year =          {2013},
  doi =           {10.1093/mnras/stt1163},
  issn =          {0035-8711},
}

@article{wiener2017CosmicRaydrivenGalactic,
  author =        {Wiener, Joshua and Pfrommer, Christoph and
                   Oh, S. Peng},
  journal =       {Monthly Notices of the Royal Astronomical Society},
  month =         jan,
  pages =         {stx127},
  title =         {Cosmic Ray-Driven Galactic Winds: Streaming or
                   Diffusion?},
  year =          {2017},
  doi =           {10.1093/mnras/stx127},
  issn =          {0035-8711, 1365-2966},
}

@article{aloisio2015NonlinearCosmicRay,
  author =        {Aloisio, R. and Blasi, P. and Serpico, P. D.},
  journal =       {Astronomy and Astrophysics},
  month =         nov,
  pages =         {A95},
  title =         {Nonlinear Cosmic Ray {{Galactic}} Transport in the
                   Light of {{AMS-02}} and {{Voyager}} Data},
  volume =        {583},
  year =          {2015},
  doi =           {10.1051/0004-6361/201526877},
  issn =          {0004-6361},
}

@article{evoli2018,
  author =        {Evoli, Carmelo and Linden, Tim and Morlino, Giovanni},
  journal =       {Physical Review D},
  month =         sep,
  pages =         {063017},
  publisher =     {APS},
  title =         {Self-Generated Cosmic-Ray Confinement in {{TeV}}
                   Halos: {{Implications}} for {{TeV}} Gamma -Ray
                   Emission and the Positron Excess},
  volume =        {98},
  year =          {2018},
  doi =           {10.1103/PhysRevD.98.063017},
  issn =          {1550-79980556-2821},
}

@article{thomas2019CosmicrayHydrodynamicsAlfvenwave,
  author =        {Thomas, T. and Pfrommer, C.},
  journal =       {Monthly Notices of the Royal Astronomical Society},
  month =         may,
  number =        {3},
  pages =         {2977--3008},
  publisher =     {Oxford Academic},
  title =         {Cosmic-Ray Hydrodynamics: {{Alfv\'en-wave}} Regulated
                   Transport of Cosmic Rays},
  volume =        {485},
  year =          {2019},
  doi =           {10.1093/mnras/stz263},
  issn =          {0035-8711},
}

@article{hopkins2021TestingPhysicalModels,
  author =        {Hopkins, Philip F. and Squire, Jonathan and
                   Chan, T. K. and Quataert, Eliot and Ji, Suoqing and
                   Kere{\v s}, Du{\v s}an and
                   {Faucher-Gigu{\`e}re}, Claude-Andr{\'e}},
  journal =       {Monthly Notices of the Royal Astronomical Society},
  month =         mar,
  pages =         {4184--4213},
  publisher =     {OUP},
  title =         {Testing Physical Models for Cosmic Ray Transport
                   Coefficients on Galactic Scales: Self-Confinement and
                   Extrinsic Turbulence at {$\sim$}{{GeV}} Energies},
  volume =        {501},
  year =          {2021},
  doi =           {10.1093/mnras/staa3691},
  issn =          {0035-8711},
}

@phdthesis{holcombMicrophysicsGyroresonantStreaming2019,
  author =        {Holcomb, Cole James},
  journal =       {Ph.D. Thesis},
  month =         jan,
  school =        {Princeton University},
  title =         {The {{Microphysics}} of {{Gyroresonant Streaming
                   Instabilities}} and {{Cosmic Ray Self-Confinement}}},
  year =          {2019},
}

@article{wentzel1968HydromagneticWavesExcited,
  author =        {Wentzel, Donat G.},
  journal =       {The Astrophysical Journal},
  month =         jun,
  pages =         {987},
  title =         {Hydromagnetic {{Waves Excited}} by {{Slowly Streaming
                   Cosmic Rays}}},
  volume =        {152},
  year =          {1968},
  doi =           {10.1086/149611},
  issn =          {0004-637X, 1538-4357},
}

@article{zweibel2003,
  author =        {Zweibel, Ellen G.},
  journal =       {The Astrophysical Journal},
  month =         apr,
  number =        {2},
  pages =         {625--637},
  title =         {Cosmic-{{Ray History}} and {{Its Implications}} for
                   {{Galactic Magnetic Fields}}},
  volume =        {587},
  year =          {2003},
  doi =           {10.1086/368256},
  issn =          {0004-637X, 1538-4357},
}

@article{lemmerz2024CouplingMultifluidDynamics,
  author =        {Lemmerz, Rouven and Shalaby, Mohamad and
                   Thomas, Timon and Pfrommer, Christoph},
  journal =       {Journal of Plasma Physics},
  month =         feb,
  number =        {1},
  pages =         {905900104},
  title =         {Coupling Multi-Fluid Dynamics Equipped with
                   {{Landau}} Closures to the Particle-in-Cell Method},
  volume =        {90},
  year =          {2024},
  doi =           {10.1017/S0022377823001113},
  issn =          {0022-3778, 1469-7807},
}

@article{hunana2018NewClosuresMore,
  author =        {Hunana, P. and Zank, G. P. and Laurenza, M. and
                   Tenerani, A. and Webb, G. M. and Goldstein, M. L. and
                   Velli, M. and Adhikari, L.},
  journal =       {Physical Review Letters},
  month =         sep,
  number =        {13},
  pages =         {135101},
  publisher =     {American Physical Society},
  title =         {New {{Closures}} for {{More Precise Modeling}} of
                   {{Landau Damping}} in the {{Fluid Framework}}},
  volume =        {121},
  year =          {2018},
  doi =           {10.1103/PhysRevLett.121.135101},
}

@article{lagageCesarsky1983,
  author =        {Lagage, P. O. and Cesarsky, C. J.},
  journal =       {Astronomy and Astrophysics},
  month =         sep,
  pages =         {249--257},
  title =         {The Maximum Energy of Cosmic Rays Accelerated by
                   Supernova Shocks.},
  volume =        {125},
  year =          {1983},
  issn =          {0004-6361},
}

@article{schroer2025,
  title = {Role of {{Nonlinear Landau Damping}} for {{Cosmic-Ray Transport}}},
  author = {Schroer, Benedikt and Caprioli, Damiano and Blasi, Pasquale},
  year = 2025,
  month = jan,
  journal = {Physical Review Letters},
  volume = {134},
  number = {4},
  pages = {045201},
  issn = {0031-9007, 1079-7114},
  doi = {10.1103/PhysRevLett.134.045201},
  urldate = {2026-10-09},
  langid = {english}
}

@article{schroer2026,
  title = {Investigating Nonlinear {{Landau}} Damping in Hybrid Simulations},
  author = {Schroer, Benedikt and Caprioli, Damiano and Blasi, Pasquale},
  year = 2026,
  month = jan,
  journal = {Physical Review D},
  volume = {113},
  pages = {023025},
  publisher = {APS},
  issn = {1550-79980556-2821},
  doi = {10.1103/cllv-6m95},
  urldate = {2026-10-09}
}

@article{armillotta2021,
  title = {Cosmic-{{Ray Transport}} in {{Simulations}} of {{Star}}-Forming {{Galactic Disks}}},
  author = {Armillotta, Lucia and Ostriker, Eve C. and Jiang, Yan-Fei},
  year = 2021,
  month = nov,
  journal = {The Astrophysical Journal},
  volume = {922},
  pages = {11},
  publisher = {IOP},
  issn = {0004-637X},
  doi = {10.3847/1538-4357/ac1db2},
  urldate = {2026-10-12}
}

@article{weidl2019BackgroundgyroresonantCharacterBells,
  title = {On the {{Background}}-Gyroresonant {{Character}} of {{Bell}}'s {{Instability}} in the {{Large}}-Current {{Regime}}},
  author = {Weidl, Martin S. and Winske, Dan and Niemann, Christoph},
  year = 2019,
  month = feb,
  journal = {The Astrophysical Journal},
  volume = {872},
  pages = {48},
  publisher = {IOP},
  issn = {0004-637X},
  doi = {10.3847/1538-4357/aafad0},
  urldate = {2024-10-24}
}

@article{mattor1992CanLandaufluidModels,
  author =        {Mattor, Nathan},
  journal =       {Physics of Fluids B: Plasma Physics},
  month =         dec,
  number =        {12},
  pages =         {3952--3961},
  title =         {Can {{Landau}}-fluid Models Describe Nonlinear
                   {{Landau}} Damping?},
  volume =        {4},
  year =          {1992},
  doi =           {10.1063/1.860350},
  issn =          {0899-8221},
}

@article{medvedev1996,
  author =        {Medvedev, M. V. and Diamond, P. H.},
  journal =       {Physics of Plasmas},
  month =         mar,
  number =        {3},
  pages =         {863--873},
  title =         {Fluid {{Models}} for {{Kinetic Effects}} on
                   {{Coherent Nonlinear Alfven Waves}}. {{I}}.
                   {{Fundamental Theory}}},
  volume =        {3},
  year =          {1996},
  doi =           {10.1063/1.871790},
  issn =          {1070-664X, 1089-7674},
}

@article{medvedev1997,
  author =        {Medvedev, M. V. and Shevchenko, V. I. and
                   Diamond, P. H. and Galinsky, V. L.},
  journal =       {Physics of Plasmas},
  month =         may,
  number =        {5},
  pages =         {1257--1285},
  title =         {Fluid {{Models}} for {{Kinetic Effects}} on
                   {{Coherent Nonlinear Alfven Waves}}. {{II}}.
                   {{Numerical Solutions}}},
  volume =        {4},
  year =          {1997},
  doi =           {10.1063/1.872356},
  issn =          {1070-664X, 1089-7674},
}

@article{lemmerz2025,
  author =        {Lemmerz, Rouven and Shalaby, Mohamad and
                   Pfrommer, Christoph and Thomas, Timon},
  journal =       {The Astrophysical Journal},
  month =         jan,
  number =        {1},
  pages =         {34},
  publisher =     {The American Astronomical Society},
  title =         {The {{Theory}} of {{Resonant Cosmic Ray}}--Driven
                   {{Instabilities}}---{{Growth}} and {{Saturation}} of
                   {{Single Modes}}},
  volume =        {979},
  year =          {2025},
  doi =           {10.3847/1538-4357/ad8eb3},
  issn =          {0004-637X},
}

@article{hahm2015IonHeatingNonlinear,
  author =        {Hahm, T. S.},
  journal =       {Plasma Science and Technology},
  month =         jul,
  number =        {7},
  pages =         {534--538},
  title =         {Ion {{Heating}} from {{Nonlinear Landau Damping}} of
                   {{High Mode Number Toroidal Alfv\'en Eigenmodes}}},
  volume =        {17},
  year =          {2015},
  doi =           {10.1088/1009-0630/17/7/02},
  issn =          {1009-0630},
}

@article{armstrong1995,
  author =        {Armstrong, J. W. and Rickett, B. J. and
                   Spangler, S. R.},
  journal =       {The Astrophysical Journal},
  month =         apr,
  pages =         {209},
  publisher =     {IOP},
  title =         {Electron {{Density Power Spectrum}} in the {{Local
                   Interstellar Medium}}},
  volume =        {443},
  year =          {1995},
  doi =           {10.1086/175515},
  issn =          {0004-637X},
}

@article{chepurnov2010,
  author =        {Chepurnov, A. and Lazarian, A.},
  journal =       {The Astrophysical Journal},
  month =         feb,
  pages =         {853--858},
  publisher =     {IOP},
  title =         {Extending the {{Big Power Law}} in the {{Sky}} with
                   {{Turbulence Spectra}} from {{Wisconsin H$\alpha$
                   Mapper Data}}},
  volume =        {710},
  year =          {2010},
  doi =           {10.1088/0004-637X/710/1/853},
  issn =          {0004-637X},
}

@article{farmer2004WaveDampingMagnetohydrodynamic,
  author =        {Farmer, Alison J. and Goldreich, Peter},
  journal =       {The Astrophysical Journal},
  month =         apr,
  number =        {2},
  pages =         {671--674},
  title =         {Wave {{Damping}} by {{Magnetohydrodynamic
                   Turbulence}} and {{Its Effect}} on {{Cosmic}}-{{Ray
                   Propagation}} in the {{Interstellar Medium}}},
  volume =        {604},
  year =          {2004},
  doi =           {10.1086/382040},
  issn =          {0004-637X, 1538-4357},
}

@article{yan2002ScatteringCosmicRays,
  author =        {Yan, Huirong and Lazarian, A.},
  journal =       {Physical Review Letters},
  month =         dec,
  pages =         {281102},
  publisher =     {APS},
  title =         {Scattering of {{Cosmic Rays}} by
                   {{Magnetohydrodynamic Interstellar Turbulence}}},
  volume =        {89},
  year =          {2002},
  doi =           {10.1103/PhysRevLett.89.281102},
  issn =          {0031-9007},
}

@article{lazarian2021DiffusionCosmicRays,
  author =        {Lazarian, Alex and Xu, Siyao},
  journal =       {The Astrophysical Journal},
  month =         dec,
  number =        {1},
  pages =         {53},
  title =         {Diffusion of {{Cosmic Rays}} in {{MHD Turbulence}}
                   with {{Magnetic Mirrors}}},
  volume =        {923},
  year =          {2021},
  doi =           {10.3847/1538-4357/ac2de9},
  issn =          {0004-637X, 1538-4357},
}

@article{cerri2024,
  author =        {Cerri, Silvio Sergio},
  journal =       {Astronomy \& Astrophysics},
  month =         aug,
  pages =         {A182},
  title =         {Revisiting the Role of Cosmic-Ray Driven {{Alfv\'en}}
                   Waves in Pre-Existing Magnetohydrodynamic Turbulence:
                   {{I}}. {{Turbulent}} Damping Rates and Feedback on
                   Background Fluctuations},
  volume =        {688},
  year =          {2024},
  doi =           {10.1051/0004-6361/202449492},
  issn =          {0004-6361, 1432-0746},
}

@article{shalabyNewCosmicRaydriven2020,
  author =        {Shalaby, Mohamad and Thomas, Timon and
                   Pfrommer, Christoph},
  journal =       {The Astrophysical Journal},
  month =         feb,
  number =        {2},
  pages =         {206},
  title =         {A {{New Cosmic-Ray-driven Instability}}},
  volume =        {908},
  year =          {2021},
  doi =           {10.3847/1538-4357/abd02d},
  issn =          {1538-4357},
}

@techreport{kulsrud1978,
  author =        {Kulsrud, R.M.},
  month =         jul,
  number =        {PPPL-1459, 6879685},
  pages =         {PPPL-1459, 6879685},
  title =         {Propagation of Cosmic Rays through a Plasma},
  year =          {1978},
  doi =           {10.2172/6879685},
}

@article{bai2019,
  author =        {Bai, Xue-Ning and Ostriker, Eve C. and
                   Plotnikov, Illya and Stone, James M.},
  journal =       {The Astrophysical Journal},
  month =         may,
  number =        {1},
  pages =         {60},
  publisher =     {The American Astronomical Society},
  title =         {Magnetohydrodynamic {{Particle-in-cell Simulations}}
                   of the {{Cosmic-Ray Streaming Instability}}: {{Linear
                   Growth}} and {{Quasi-linear Evolution}}},
  volume =        {876},
  year =          {2019},
  doi =           {10.3847/1538-4357/ab1648},
  issn =          {0004-637X},
}

@article{denton1995DfAlgorithm,
  author =        {Denton, Richard E. and Kotschenreuther, M.},
  journal =       {Journal of Computational Physics},
  month =         jul,
  number =        {2},
  pages =         {283--294},
  title =         {{$\delta$}f {{Algorithm}}},
  volume =        {119},
  year =          {1995},
  doi =           {10.1006/jcph.1995.1136},
  issn =          {0021-9991},
}

@article{kunz2014,
  author =        {Kunz, Matthew W. and Stone, James M. and
                   Bai, Xue-Ning},
  journal =       {Journal of Computational Physics},
  month =         feb,
  pages =         {154--174},
  title =         {Pegasus: {{A}} New Hybrid-Kinetic Particle-in-Cell
                   Code for Astrophysical Plasma Dynamics},
  volume =        {259},
  year =          {2014},
  doi =           {10.1016/j.jcp.2013.11.035},
  issn =          {0021-9991},
}

@article{sudan1971TheoryTriggeredVLF,
  author =        {Sudan, R. N. and Ott, E.},
  journal =       {Journal of Geophysical Research},
  month =         jul,
  number =        {19},
  pages =         {4463--4476},
  title =         {Theory of Triggered {{VLF}} Emissions},
  volume =        {76},
  year =          {1971},
  doi =           {10.1029/JA076i019p04463},
  issn =          {01480227},
}

@article{otani1988ApplicationNonlinearDynamical,
  author =        {Otani, Niels F.},
  journal =       {The Physics of Fluids},
  month =         jun,
  number =        {6},
  pages =         {1456--1464},
  title =         {Application of Nonlinear Dynamical Invariants in a
                   Single Electromagnetic Wave to the Study of the
                   {{Alfv\'en}}-ion-cyclotron Instability},
  volume =        {31},
  year =          {1988},
  doi =           {10.1063/1.866736},
  issn =          {0031-9171},
}

@article{zachary1989LongtimeEvolutionLowdensity,
  author =        {Zachary, Andrew L. and Cohen, Bruce I. and
                   Max, Claire E. and Arons, Jonathan},
  journal =       {Journal of Geophysical Research: Space Physics},
  number =        {A3},
  pages =         {2443--2458},
  title =         {The Long-Time Evolution of a Low-Density Ion Beam},
  volume =        {94},
  year =          {1989},
  doi =           {10.1029/JA094iA03p02443},
  issn =          {2156-2202},
}

@article{skilling1975CosmicRayStreamingI,
  author =        {Skilling, J.},
  journal =       {Monthly Notices of the Royal Astronomical Society},
  month =         sep,
  number =        {3},
  pages =         {557--566},
  publisher =     {Oxford University Press (OUP)},
  title =         {Cosmic {{Ray Streaming--I Effect}} of {{Alfven
                   Waves}} on {{Particles}}},
  volume =        {172},
  year =          {1975},
  doi =           {10.1093/mnras/172.3.557},
  issn =          {0035-8711, 1365-2966},
}

@article{ruszkowski2017,
	title = {Cosmic-{{Ray Feedback Heating}} of the {{Intracluster Medium}}},
	author = {Ruszkowski, Mateusz and Yang, H.-Y. Karen and Reynolds, Christopher S.},
	year = 2017,
	month = jul,
	journal = {The Astrophysical Journal},
	volume = {844},
	number = {1},
	pages = {13},
	publisher = {The American Astronomical Society},
	issn = {0004-637X},
	doi = {10.3847/1538-4357/aa79f8},
	urldate = {2023-01-17},
	langid = {english}
}
\bibliographystyle{aasjournalv7.1}

\end{document}